\documentclass[10pt,preprint,longnamesfirst]{aastex}

\slugcomment{ApJ, in press: 22 September 2002}

\shorttitle{Our Sun.\ IV.\ Standard Model Uncertainties}
\shortauthors{Boothroyd \& Sackmann}

\newcommand{\iso}[2]{\hbox{${}^{#1}$#2}}
\newcommand{\gammaone}{\Gamma_{\!1}}
\newcommand{\gammatwo}{\Gamma_{\!2}}
\newcommand{\gammathree}{\Gamma_{\!3}}

\begin{document}

\title{Our Sun.\ IV. The Standard Model and Helioseismology:\\
Consequences of Uncertainties in Input Physics\\
and in Observed Solar Parameters}

\author{Arnold I. Boothroyd\altaffilmark{1} and
 I.-Juliana Sackmann\altaffilmark{2}}
\affil{W. K. Kellogg Radiation Laboratory 106-38,
 California Institute of Technology,\\
 Pasadena, CA 91125}
\email{boothroy@cita.utoronto.ca}
\email{ijs@caltech.edu}
\altaffiltext{1}{Present address: CITA, U. of Toronto, 60 St.\ George Street,
 Toronto, Ontario, Canada M5S 3H8}
\altaffiltext{2}{Present address: West Bridge Laboratory 103-33,
 California Institute of Technology, Pasadena, CA 91125}

\begin{abstract}

Helioseismology provides a powerful tool to explore the
deep interior of the Sun.  Measurements of solar interior quantities
are provided with unprecedented accuracy:
for example, the adiabatic sound speed~$c$
can be inferred with an accuracy of a few parts in~$10^4$.
This has become a serious challenge to theoretical models of the Sun.
Therefore, we have undertaken a self-consistent, systematic study of
sources of uncertainties in the standard solar model,
which must be understood before the helioseismic observations can
be used as constraints on the theory.
This paper focusses on our own current calculations, 
but is also a review paper summarizing
the latest calculations of other authors.
We find that the largest uncertainty in the sound speed~$c$ in the
solar interior, namely, 3 parts in~$10^3$,
arises from uncertainties in the observed photospheric abundances of the
elements: C, N, O, and Ne have uncertainties of~$\sim 15$\%,
leading to an uncertainty of $\sim 10$\% in the photospheric $Z/X$ ratio.
Uncertainties of 1 part in~$10^3$ in the sound speed~$c$ arise, in each
case, from
(1)~the $\sim 4$\% uncertainty in the OPAL opacities, (2)~the $\sim 5$\%
uncertainty in the basic $pp$ nuclear reaction rate, (3)~the $\sim 15$\%
uncertainty in the diffusion constants for the gravitational settling
of helium, and (4)~the $\sim 50$\% uncertainties in diffusion constants
for the heavier elements.  (Other investigators have shown that similar
uncertainties arise from uncertainties in the interior equation of state
and in rotation-induced turbulent mixing.)  In the convective envelope
{\it only}, uncertainties in~$c$ of order 1 part in~$10^3$ arise from
the uncertainty of a few parts in~$10^4$ in the solar radius, and from
uncertainties in the low-temperature equation of state.
Other current uncertainties, namely, in the solar age and
luminosity, in nuclear rates other than the
$pp$~reaction, and in the low-temperature molecular opacities,
have no significant effect on the
quantities that can be inferred from helioseismic observations.
Significant uncertainty in the convective envelope position~$R_{ce}$
(of up to 3~times the observational uncertainty
of $\pm 0.001 \; R_\odot$) arises only from uncertainties in $Z/X$,
opacities, the $pp$~rate, and helium diffusion constants;
the envelope helium
abundance~$Y_e$ is significantly affected ($\pm 0.005$) only by extreme 
variations in~$Z/X$, opacities, or diffusion constants, and is always
consistent with the ``observed'' range
of helioseismically inferred $Y_e$ values.
Our predicted pre-main-sequence solar lithium depletion
is a factor of $\sim 20$ (an order of magnitude larger than that predicted by
earlier models that neglected gravitational settling and used
older opacities), and
is uncertain by a factor of~2.  The predicted neutrino capture rate is
uncertain by~$\sim 30$\% for the \iso{37}{Cl} experiment and by~$\sim 3$\%
for the \iso{71}{Ga} experiments (not including uncertainties in the
capture cross sections), while the \iso8B neutrino flux is
uncertain by~$\sim 30$\%.
\end{abstract}



\section{Introduction} \label{sec:intro}

Helioseismology provides a powerful tool to explore the
deep interior of the Sun.  Measurements of solar interior quantities
are provided with unprecedented accuracy.  The Michelson Doppler
Imager (MDI) on the Solar and Heliospheric Observatory (SOHO) spacecraft
\citep{Rho+97},
the Global Oscillation Network Group
\citep[GONG: see, e.g.,][]{Gough+96,Harvey+96},
the Birmingham Solar Oscillation Network
\citep[BiSON:][]{Chap+96},
and the Low-{\it l} instrument
\citep[LOWL:][]{TomST95}
provide helioseismic frequency measurements with accuracies of a
few parts in~$10^5$.  From these, the sound speed~$c$ throughout
most of the solar interior can be inferred with
an accuracy of a few parts in~$10^4$, as can the adiabatic
index~$\Gamma_{\!1}$; the density can be inferred with an accuracy
of a few parts in~$10^3$
\citep{BasuPB00,BahPB01}.
The depth of the solar envelope convection can be measured with
an accuracy of nearly a part in~$10^3$
\citep{BasuA97}.
The abundance of helium in the solar envelope can also be inferred,
but this value depends on the solar equation of state
\citep[see, e.g.,][]{BasuA95,Rich+98,DiMauro+02};
as discussed in \S~\ref{ssec:helium}, this inferred helium abundance
appears to be uncertain at the level of a few percent (i.e., an
uncertainty of at least~0.005 in the helium mass fraction~$Y_e$).

The ultimate goal of our work was to explore systematically a wide range
of solar models with relatively modest amounts of mass loss on the early
main sequence, and to test their viability via helioseismological
measurements (as well other observational constraints).  This mass loss
investigation is presented in our companion paper ``Our Sun~V''
\citep{SB02}.
Since the
consequences of moderate early solar mass loss are expected to be
small, it is essential to understand the consequences of uncertainties
in the input physics and in the input parameters of the solar model.
The physics inputs include the equation of state, opacities,
nuclear rates, diffusion constants, the treatment of convection
(including the possibility of overshoot), and the effects of rotation
and mass loss.
Observed solar parameters include the solar age and the present solar
radius, luminosity, surface composition, and solar wind.
A considerable amount of work has been published investigating many
of the above effects, as will be discussed in \S~\ref{sec:results}.
However, some basic uncertainties still warrant further attention.  Before
proceeding to our mass loss work, we found it necessary to try to extend
the investigations of the above uncertainties.  In particular, the major
consequences arising from uncertainties in the present observed
solar surface Z/X ratio, in the basic \hbox{\it p-p} chain
rate, and in the diffusion constants for gravitational settling of
helium and the heavy elements,
have not been addressed sufficiently in the recent (most accurate)
work of other investigators.  To determine the consequences of these
uncertainties, we computed standard solar models with various values of
the Z/X ratio, the \hbox{\it p-p} chain rate, and the diffusion
constants (lying in the permitted
ranges).  To obtain a set of self-consistent results, we also investigated
the consequences
of a number of other uncertainties, namely, in other nuclear
rates, in envelope opacities, in the equation of
state, from different methods of
handling interior opacities, and in the solar age, luminosity, and radius.
It is the aim of this paper to present the
effects of the above uncertainties
on the run of the sound speed and density in the solar interior, and
on the radius~$R_{ce}$ of solar convection and the solar envelope
helium abundance~$Y_e$.
We also present the effects on the pre-main-sequence
solar lithium depletion (excluding
rotation effects) and on the production of solar neutrinos.  (Note that
lithium depletion in a non-rotating standard solar model occurs
entirely on the pre-main-sequence; this predicted depletion is
significantly smaller than the observed solar lithium depletion factor.
Rotational mixing on the main sequence is generally invoked to account for
the remaining lithium depletion).


\section{Methods} \label{sec:methods}


We computed a reference standard solar model using up-to-date physics plus
the observed values for the solar parameters.  Several dozen variant solar
models were also computed,
in which one of these ``inputs'' was varied within the allowed uncertainties.
(In a few cases, more than one of the ``inputs'' was varied, or the size of
the variation exceeded the size of the quoted uncertainty in order to get a
better estimate of the sensitivity.)  By comparing with
the reference standard solar model, the sensitivity to the uncertainties in
the ``inputs'' could be determined.
The stellar evolution code used to compute these solar models
is descended from that used earlier in our ``Sun~III'' paper
(\citealt{SBK93};
see also
\citealt{BS99}),
but has been extensively updated for improved accuracy, including
provision for much finer zoning as well as up-to-date input physics.

\paragraph{Equation of State:}
The reference standard solar model used the OPAL equation of state
\citep{RogSI96}
in the interior, and the MHD equation of state
\citep{Dap+88}
where it was designed to be accurate, namely, in the outer envelope at
$\log\,\rho \lesssim -2$ (this corresponds to $r \gtrsim 0.94\;R_\odot$
and $\log\,T \lesssim 5.5$
in the present Sun).  A version of the MHD equation of state computation
program was kindly provided to us
\citetext{D.~Mihalas 1999, private communication},
and minor modifications allowed computation of the MHD equation of state
for various hydrogen abundances and metallicities down to pre-main-sequence
photospheric temperatures and even below
\citetext{A.~I.~Boothroyd, in preparation}.
In both cases, the equation of state was interpolated in metallicity~$Z$ as
well as in hydrogen abundance~$X$, temperature~$T$, and density~$\rho$,
in order to take into account metallicity variations due to diffusion
and nuclear burning.  Tables for the MHD equation of state were
computed on the same $\{X,Z,T,\rho\}$ grid as
the OPAL tables, with the following exceptions: for MHD, the upper 
$\rho$~limit was slightly higher than for OPAL at low temperatures, but lower
at high temperatures (since it never exceeded $\log\,\rho = -1.5$); below
$log\,T = 3.7$, intervals of $\Delta\,\log\,T = 0.01$ were used for the MHD 
$T$-grid, the same as the average $\log\,T$ interval in the OPAL grid for
$0.005 \le T_6 \le 0.006$ ($3.7 \lesssim \log\,T \lesssim 3.78$); and for
$0.006 \le T_6 \le 0.05$ ($3.78 \lesssim \log\,T \lesssim 4.7$), extra
$T$-grid points were added to double the density of the grid as a function
of temperature in this region.  By computing some MHD tables at intermediate 
grid-values, the interpolation accuracy in each of $\{X,Z,T,\rho\}$ was
tested.  In the outer solar convective zone ($r \gtrsim 0.95 \; R_\odot$),
the fractional error in
the pressure~$P$ was $\lesssim 2 \times 10^{-4}$, and for derivative
quantities such as $C_v$, $\chi_T$, $\chi_\rho$, and~$\gammaone$, the
fractional error was $\lesssim 2 \times 10^{-3}$; deeper in the Sun
($r < 0.95 \; R_\odot$), the interpolation errors were nearly an order
of magnitude smaller.  The interpolation accuracy is thus several times
better than the quoted precision of the OPAL equation of state in the
corresponding regions
\citep{RogSI96}.

Variant cases were tested where the OPAL equation of state was used down
to $\log\,T = 4$ or even all the
way down to its lower limit of validity at $\log\,T = 3.7$ --- in the latter
case, the MHD equation of state was relevant only to the pre-main-sequence
evolution.  The MHD equation of state tables computed for this work
considered H and~He, plus a 13-element subset of the
\citet{GreN93}
heavy element composition (i.e., C, N, O, Ne, Na, Mg, Al, Si, S, Ar, K, Ca,
and~Fe); besides the effects due to neutrals and ions,
effects due to H$^-$, H$_2$, and~H$_2^+$ were accurately taken
into account, and approximate effects of the molecules C$_2$, N$_2$, O$_2$,
CH, CN, CO, NH, NO, OH, CO$_2$, and~H$_2$O (although these molecules have
no effect for solar models: they were actually added in anticipation of
use in asymptotic giant branch models).  The OPAL equation of state
considers only H, He, C, N, O, and~Ne (plus hydrogen-molecule effects);
however, the equation of state is quite insensitive to the precise makeup
of the metallicity, and the OPAL equation of state may well be the more
accurate one in the region where both are valid
\citep{RogSI96,GongDN01}.

It has been noted that patching together equations of state can introduce
spurious effects
\citep{Dap+93,BasuDN99}.
Such spurious effects were minimized in the present work by (1)~performing a
gradual switchover over a finite density or temperature region, to avoid any
discontinuities, and (2)~choosing switchover regions where the two equations
of state were quite similar.  The three switchover regions comprised
$-1.5 > \log\,\rho > -2$ (corresponding to $5.8 > \log\,T > 5.5$ in the 
present Sun), $4.0 > \log\,T > 3.9$, or $3.75 > \log\,T > 3.7$; in the first
and last of these regions, the MHD and OPAL equations of state under solar
conditions differ typically by several parts in~$10^3$ (with differences
in the pressure~$P$ being several times smaller than this), i.e., only
a few times worse than the expected precision of the equations of state.
Near $\log\,T = 4$, the differences were nearly an order of magnitude smaller
(and also remained small over a very wide range of densities).  Any spurious
effects from the switchover should be smaller than spurious effects arising
from the fact that the OPAL equation of state is not self-consistent, with
the size of the inconsistency (in the outer Solar envelope)
being comparable to the size of the differences
between the OPAL and MHD equations of state
(as discussed in \S~\ref{ssec:profiles}).

\paragraph{Opacities:}
For interior temperatures ($\log\,T > 4$),
our reference standard solar model used the 1995 OPAL opacities
\citep{IglR96}
--- these opacities use the
\citet{GreN93}
``GN93'' solar composition, and we refer to these opacities as
``$\kappa_{\rm OPAL:GN93}$.''
The online opacity computation feature of the OPAL web page\footnote{
\url[http://www-phys.llnl.gov/Research/OPAL/]{http://www-phys.llnl.gov/Research/OPAL/}}
allowed computation of OPAL opacities appropriate to the more recent
\citet{GreS98}
``GS98'' mixture
(``$\kappa_{\rm OPAL:GS98}$''); these were tested in variant models,
as were opacities appropriate to the older
\citet{Gre84}
``Gr84'' mixture (``$\kappa_{\rm OPAL:Gr84}$'').
The even older and much less accurate Los Alamos (LAOL) opacities
``$\kappa_{\rm LAOL85}$''
\citetext{Keady 1985, private communication}
were also tested in one case.
At cool envelope temperatures ($\log\,T \lesssim 4$),
our reference standard solar model used the
\citet{AlexF94}
opacities ``$\kappa_{\rm Alexander}$'' (which include molecular opacities).
We also tested the effect of using the
\citet{Shar92}
molecular opacities ``$\kappa_{\rm Sharp}$''
at cool envelope temperatures instead.

In our reference standard solar model, we did our best to account for
temporal and spatial variations in the opacity due to composition
changes from diffusion and nuclear burning.  The OPAL opacity tables
allow interpolation of the opacity as a function of the hydrogen
abundance and the metallicity~$Z$; the abundances of the metals
comprising~$Z$ are always proportional to~$Z$ in the OPAL tables,
i.e., a ``scaled solar'' distribution.  In addition, the OPAL
opacity tables contain mixtures with excess carbon and oxygen (beyond that
contained in the scaled solar metallicity), allowing interpolation
in carbon and oxygen abundances.

To take into account the variations in
abundance due to diffusion, the metallicity value~$Z_\kappa$ that
we used for metallicity interpolation in the OPAL opacity tables
was proportional to the abundances of the elements heavier than oxygen:
$Z_\kappa = Z_h$, where
$Z_h \equiv Z_0 [ \sum_{heavy} X_i ] / [ \sum_{heavy} (X_i)_0 ]$,
where $Z_0$ and $(X_i)_0$ are the protosolar metallicity and composition,
respectively, and $\sum_{heavy}$ refers to a sum over elements heavier
than oxygen.  In other words,
we scaled the initial solar metallicity by the shift in heavy element
abundances resulting from diffusion (note that our diffusion routines
assumed that all metals diffused alike).  However, for the CNO elements, there
are additional changes due to nuclear burning, so that the CNO abundance
profiles are not proportional to the heavy element abundance profiles;
\citet{Turc+98}
find that conversion of C and~O into~N results in an opacity change
of~$\sim 1$\% that cannot be modelled by a change in~$Z$ alone.
These variations in the CNO elements relative to~$Z_h$ were accounted for
in an approximate manner by an additional two-dimensional interpolation
in nominal ``excess carbon and oxygen'' abundances ${\rm C}_{ex}$
and~${\rm O}_{ex}$, where these excess abundances account for
any variation in the CNO abundances relative to the scaled solar
metallicity~$Z_\kappa$ of the OPAL opacity tables.
Since there were no explicit opacity tables
for changes in nitrogen, the best one could do was to distribute
excess nitrogen equally between ${\rm C}_{ex}$ and~${\rm O}_{ex}$
(i.e., to assume that nitrogen opacities were midway between those
of carbon and oxygen):
${\rm C}_{ex} = {\rm C - C_0 \, Z_\kappa / Z_0
 + 0.5 ( N - N_0 \, Z_\kappa / Z_0 )}$ and
${\rm O}_{ex} = {\rm O - O_0 \, Z_\kappa / Z_0
 + 0.5 ( N - N_0 \, Z_\kappa / Z_0 )}$,
such that ${\rm CO}_{ex} \equiv {\rm C}_{ex} + {\rm O}_{ex} = Z - Z_\kappa$.

Note that in general, as first carbon and then oxygen is burned to nitrogen,
either ${\rm C}_{ex}$ or~${\rm O}_{ex}$ will be negative, implying
{\it extrapolation\/} of the OPAL tables in the direction of zero C or~O
by a non-negligible fraction of~$Z$.  One might consider the uncertainties
in such an extrapolation to be worse than the error inherent in treating
all CNO opacities alike; in this case, if one of ${\rm C}_{ex}$
or~${\rm O}_{ex}$ was negative, one would set it to zero and subtract
an equivalent amount from the other (so that the sum
${\rm C}_{ex} + {\rm O}_{ex}$ remains unchanged) --- this was the method
followed in
our reference models.  However, we also tested the case where negative
values were allowed, extrapolating as required.  These turned out to
yield essentially identical results to the reference model: the rms
difference was only~0.00006 in the sound speed profile $\Delta c / c$
and~0.0003 in $\Delta \rho / \rho$, comparable to the estimated
numerical accuracy of the models.
An alternative case where this latter CNO variation was not approximated
by excess~CO at all, i.e., having
$Z_\kappa = Z_h$ but CO$_{ex} = 0.0$, also yielded almost
identical results (rms difference of~0.00005 in $\Delta c / c$
and~0.0004 in $\Delta \rho / \rho$).  
Similarly negligible rms differences were found for several
cases testing different prescriptions for
defining~$Z_h$, ${\rm C}_{ex}$, and~${\rm O}_{ex}$
(differing in whether mass fractions or number densities were used,
in how the excess nitrogen was divided between ${\rm C}_{ex}$
and~${\rm O}_{ex}$, and in whether negative values were allowed for
${\rm C}_{ex}$ and~${\rm O}_{ex}$).

Finally, one run was computed where CNO-interpolation in the opacities was
actually performed.  In addition to the standard OPAL opacity table,
the OPAL web page\footnote{
\url[http://www-phys.llnl.gov/Research/OPAL/type1inp.html]{http://www-phys.llnl.gov/Research/OPAL/type1inp.html}}
was used to compute an opacity table with a composition where C had been
converted to~N, and one where both C and~O had been converted to~N; the
model CNO abundances were used to interpolate in opacity among these tables.
This ``CNO-interpolation'' model also yielded results essentially identical to
the reference model (an rms difference of~0.00006 in $\Delta c / c$
and~0.0005 in $\Delta \rho / \rho$).

An estimate for the upper limit of the effects of opacity uncertainties
was made by simply setting $Z_\kappa = f \, Z$ for a constant
factor~$f = 0.9$, 0.95, 1.0, 1.05, or~1.10; note that the case $f = 1.0$
is only slightly different from the reference standard solar model.

\paragraph{Nuclear Reaction Rates:}
The reference standard solar model used the NACRE nuclear reaction rates
\citep{Ang+99},
supplemented by the \iso7{Be} electron capture rates of
\citet{GruB97}
for $\log\,T \ge 6$ and of
\citet{BahM69}
for $\log\,T < 6$ (note that this latter low-temperature region is irrelevant
for solar models).  Variant models tested cases with nuclear rates
changed according to the upper or lower limits quoted in the NACRE
compilation; one case tested the use of the older
\citet{CF88}
nuclear rates.  The program uses the minimum of the weak
\citep{Sal55},
intermediate
\citep{Grab+73},
or strong
\citep{Itoh+79,IchU83}
screening factors (note that both weak and intermediate screening formulae
include a term to take partial electron degeneracy into account) --- for solar
conditions, this choice leads to the use of weak screening, which is a very
good approximation to the exact quantum mechanical solution there
\citep[see, e.g.,][]{BahCK98,GruB98}.
Deuterium was not considered separately (it was assumed to have been
entirely burned to \iso3{He} on the early pre-main-sequence), but all
the other 15 stable isotopes up to and including \iso{18}O were considered
in detail (i.e., nuclear equilibrium was not assumed for any of them).
The other stable isotopes up to~\iso{28}{Si} were included in the code
(plus a few long-lived unstable isotopes, as well as~Fe and a category for
the sum of the other elements heavier than~Si), but their nuclear reactions
were not included since there are no significant effects under solar
conditions (except for~\iso{19}F, which was assumed to be in CNO-cycle
nuclear equilibrium for nuclear rate purposes).
Neutrino capture cross sections were taken from
\citet{BahU88},
except for the \iso8B-neutrino cross section for capture on \iso{37}{Cl},
where the more recent value (5\%~higher) of
\citet{Aufder+94}
was used.

\paragraph{Diffusion:}
A set of subroutines\footnote{
These subroutines are also available from Bahcall's web page:
\url[http://www.sns.ias.edu/~jnb/]{http://www.sns.ias.edu/${}^{\sim}$jnb/}}
were kindly provided to us
\citetext{M.~H.~Pinsonneault 1999, private communication}
that take into account the diffusion (gravitational settling) of helium and
heavy elements relative to hydrogen
\citep[see also][]{ThoulBL94,BahPW95}.
These subroutines assume that all heavy elements diffuse at the same rate
as fully-ionized~Fe; this yields surprisingly accurate results, as
may be seen by comparing to the results of a more detailed treatment
\citep{TurcC98,Turc+98}.
The upper limit of the effects of uncertainties in the
diffusion constants was estimated by simply multiplying the diffusion
constants, either for helium or for the heavy elements, by a constant
factor.

\paragraph{Convection:}
The Schwarzschild criterion was used to define convective boundaries; no
core overshooting or envelope undershooting was allowed.  Note that
\citet{MorPB97}
found that including convective core overshooting had a negligible
effect on the solar sound speed and density profiles, but on the other
hand that including convective envelope undershooting by even a tenth
of a pressure scale height moved the solar convective envelope boundary
inwards by eight times the uncertainty in the observed value, yielding
a sharp spike in the difference between observed and calculated sound
speed profiles.  Rotation-induced mixing was not considered in our
models; the effect that it would have is discussed in~\S~\ref{ssec:profiles}.

\paragraph{Composition:}
The reference standard solar model used a value of $Z/X = 0.0245$ for
the the present solar surface metals-to-hydrogen ratio (by mass fraction),
as given by
\citet{GreN93}.
Variant models tested the $\sim 13$\% higher older value of $Z/X = 0.0277$
\citep{Gre84,AndG89},
as well as the 6\% lower value of $Z/X = 0.023$ recommended as
a ``preliminary'' value by the recent work of
\citet{GreS98}.
Values of $Z/X = 0.0203$ and~0.0257 were also used, for cases where the
C, N, O, and~Ne abundances of the
\citet{GreS98}
mixture were all either decreased or increased, respectively,
by their quoted uncertainties.

For a given solar model, the presolar abundances of the heavy
elements were always taken
from the OPAL opacity tables that we used for that solar model.
For the reference standard solar model and most variants, the
``$\kappa_{\rm OPAL:GN93}$'' opacities were used, which have
the composition mix of
\citet{GreN93}.
A variant model with a low value $Z/X = 0.023$ used instead
the composition mix of
\citet{GreS98},
and the corresponding ``$\kappa_{\rm OPAL:GS98}$'' opacities.
Similarly, a variant model with a high value of
$Z/X = 0.0277$ used of the abundance mix of
\citet{Gre84},
and the corresponding ``$\kappa_{\rm OPAL:Gr84}$'' opacities.
Tests were also made with C, N, O, and Ne abundances increased or
decreased by their uncertainties of~15\% relative to~Fe,
with the OPAL opacities appropriate to
these revised mixes: we refer to these as
``$\kappa_{\rm OPAL:GN93\uparrow C-Ne}$'' and
``$\kappa_{\rm OPAL:GN93\downarrow C-Ne}$,'' respectively,
when variations are relative to the
\citet{GreN93}
mix, and as
``$\kappa_{\rm OPAL:GS98\uparrow C-Ne}$'' and
``$\kappa_{\rm OPAL:GS98\downarrow C-Ne}$''
when variations are relative to the
\citet{GreS98}
mix.

\paragraph{Solar Mass:}
A present solar mass of $M_\odot = 1.9891 \times 10^{33}$~g
\citep{CohT86}
was used in all cases.  Their quoted
uncertainty of~0.02\% is too small to have any
significant effect, and is in fact smaller than the amount of mass
lost by the Sun in the form of radiation
($\Delta M = \Delta E / c^2$ yields a mass loss
of~0.03\%, where $\Delta E$ is the total energy radiated away via
photons and neutrinos since the Sun formed).  Mass loss from the present
solar wind is also small, and even if the average solar wind over the
past 4.6~Gyr had been an order of magnitude higher than its present value,
as is suggested by measurements of noble gas isotopes in lunar rocks
\citep{Gei73,GeiB91,Ker+91},
the total amount of mass lost would be about~0.2\% (i.e., a total mass loss
of $\le 0.002 \; M_\odot$ during the Sun's lifetime up to the present).
Mass loss of this amount would yield negligible changes in the solar
sound speed profile (about a part in~$10^4$), as shown in
our companion paper ``Our Sun~V''
\citep{SB02}.
Solar mass loss was therefore ignored for all cases considered in this paper.

Note that, based on a correlation of X-ray flux with
their measured mass loss rates in nine GK~dwarfs,
\citet{Wood+02}
have recently proposed a mass loss time dependence
$\dot M \propto t^{-2.00\pm0.52}$ in such stars (i.e., in
main sequence stars with masses not too far from that of the Sun),
with a maximum mass loss rate of $\sim 10^3$ times that of the
present solar wind.  This is discussed in our companion paper ``Our Sun~V''
\citep{SB02},
and will be investigated in more detail in a future work
\citetext{A.~I.~Boothroyd \& \hbox{I.-J.}~Sackmann, in preparation}.
In summary, total solar mass loss from the formula of
\citet{Wood+02}
could be of order~$0.01\;M_\odot$, but most of this would take place
very early on the main sequence due to the $t^{-2}$ time dependence,
so the effect should be relatively small, at most a few parts in~$10^4$
in the sound speed profile.

\paragraph{Solar Luminosity:}
For the reference standard solar model and most variant cases, a present
solar luminosity of $L_\odot = 3.854 \times 10^{33}$~erg~s$^{-1}$ was
used, as discussed in
\citet{SBK93}.
A value 0.3\% lower ($3.842 \times 10^{33}$~erg~s$^{-1}$) with an
estimated 1-$\sigma$ uncertainty of~0.4\% was recently obtained by
\citet{BahPB01},
based on the observations of
\citet{FroL98}
and
\citet{Crom+96}.
Variant models considered the effect of using this more recent solar
luminosity value, and high and low values 2-$\sigma$ (0.8\%) above and
below it.  Note that most of the uncertainty in~$L_\odot$ comes not from
uncertainties in the present solar irradiance, but rather from
uncertainties in the slight long-term variability of the solar
luminosity.  For example,
\citet{Lean00}
estimates a difference of 0.2\% between the present value and that
during the seventeenth century Maunder Minimum.

\paragraph{Solar Radius:}
The reference standard solar model used a solar radius at the photosphere
($\tau = 2/3$) of $R_\odot = 695.98$~Mm
\citep{UlrR83,Gue+92}.
Variant models considered the effect of using the value of 695.78~Mm
suggested by the helioseismic $f$-mode study of
\citet{Antia98},
or the value of 695.508~Mm suggested by the solar-meridian transit study of
\citet{BroC98}.

\paragraph{Solar Age:}
In his Appendix to
\citet{BahPW95},
G.~J.~Wasserburg provides a systematic analysis of the upper and lower
bounds on the age of the Sun, as obtained from isotopic ratios measured
in meteorites.  We briefly paraphrase his discussion in this paragraph:
The protosolar nebula (out of which the Sun and the
meteorites formed) contained not only the stable isotope~\iso{27}{Al}
but also the unstable isotope~\iso{26}{Al}, which decays into~\iso{26}{Mg}
with a half-life of only 0.7~Myr.  This~\iso{26}{Al} must have been
injected into the protosolar nebula from the stellar source where it
was created.  Isotopic measurements of meteoritic
crystallized refractory condensates show that they had a ratio
$\iso{26}{Al} / \iso{27}{Al} = 5 \times 10^{-5}$ at the time they formed.
Even if the stellar source of the~\iso{26}{Al} had a very high ratio
$\iso{26}{Al} / \iso{27}{Al} \sim 1$, the time interval between the
formation of the~\iso{26}{Al} and the formation of the meteorite
cannot have exceeded $\sim 11$~Myr (due to the short decay timescale
of~\iso{26}{Al}).
The Sun cannot have formed later than these meteorites.
However, the Sun/meteorite system must have formed after the injection
of~\iso{26}{Al} into the protosolar nebula, so the Sun cannot
have formed earlier than $\sim 11$~Myr before the formation of
these meteorites.
The age of the meteorites has been accurately measured using
\iso{207}{Pb}/\iso{206}{Pb} ratios, to be $4.565 \pm 0.005$~Gyr.
It follows that the Sun cannot have formed earlier than 4.59~Gyr ago, nor
later than 4.55~Gyr ago (i.e., $4.565 + 3 \times 0.005 + 0.011$~Gyr, or
$4.565 - 3 \times 0.005$~Gyr, at the 3-$\sigma$ level).  This total solar
age estimate of, in effect, $t_\odot = 4.57 \pm 0.01$~Gyr
is in agreement with the total solar (and solar system)
age $\tau_{ss} = 4.53 \pm 0.03$~Gyr inferred previously by
\citet{Gue89}
from a formation age of $4.53 \pm 0.02$~Gyr
\citep{Wasson85}
for the meteorites and planets.

Our solar models were started relatively high on the pre-main-sequence
Hayashi track, with central temperatures below $10^6\;$K; note that the
ages~$t_\odot$ of all our models are quoted relative to this
pre-main-sequence starting point, which is within a few Myr of the
solar formation age constrained by the meteoritic ages discussed above.
A relatively high value of $t_\odot = 4.6$~Gyr was used for the reference
standard solar model (cf.\ the observationally inferred
value $t_\odot = 4.57 \pm 0.01$~Gyr from the previous
paragraph).  To get a reliable estimate of the sensitivity of
the models to the solar age uncertainty, variant models were computed
with ages differing by very large amounts, namely, $t_\odot = 4.5$
and 4.7~Gyr.

 From our Hayashi track starting point, it takes
only $\sim 3$~Myr for the luminosity on the pre-main-sequence to drop below
$1\;L_\odot$ (i.e., below the present solar luminosity), but much longer,
namely an additional 40~Myr, to reach the zero-age main sequence (ZAMS) ---
we have defined the ZAMS as the stage where the pre-main-sequence
contraction terminates and the Sun begins to expand slowly, as nuclear
burning in the core (rather than gravitational contraction) supplies
essentially all of the solar luminosity.
For the next $\sim 50$~Myr on the early main sequence, evolution is fairly
fast, as $p + {\rm C}$ reactions burn the initial carbon to nitrogen
near the Sun's center, resulting in a short-lived convective core.
Subsequently, the central carbon and nitrogen abundances approach
their CN-cycle
equilibrium values, the convective core disappears, and the Sun settles
down to burn hydrogen mainly via the $pp$-chain reactions.  Note that the
pre-main-sequence timescale implies that the {\it total solar age\/}~$t_\odot$
used in this paper can be converted into a {\it main sequence\/} solar
lifetime by subtracting about 0.04~Gyr --- this was also pointed out by
\citet{Gue89}.

\paragraph{Helioseismology:}
We compared our solar models to profiles of the solar sound
speed~$c_\odot$, density~$\rho_\odot$, and adiabatic index~$(\gammaone)_\odot$
obtained from the helioseismic reference model of
\citet{BasuPB00}\footnote{
 From the denser-grid machine-readable form of their Table~2, at
\url[http://www.sns.ias.edu/~jnb/]{http://www.sns.ias.edu/${}^{\sim}$jnb/}},
which they obtained by inversion from the helioseismic frequency
observations.  In the inversion process, a standard solar model is
required, but
\citet{BasuPB00}
demonstrated that the resulting $c_\odot$ and~$\rho_\odot$ profiles
of the helioseismic reference model are
relatively insensitive to uncertainties in the standard solar model
used for this purpose (except for uncertainties in~$R_\odot$, as
discussed in~\S~\ref{ssec:profiles}).  They estimated a net uncertainty
of few parts in~$10^4$ for the sound speed~$c_\odot$ and adiabatic
index~$(\gammaone)_\odot$, and a few parts
in~$10^3$ for the density~$\rho_\odot$.  However, in the Sun's core
($r \lesssim 0.1\;R_\odot$), systematic uncertainties in the
helioseismic sound profile are increased by a factor of~$\sim 5$; this was
demonstrated by
\citet{BahPB01},
who compared helioseismic inversions of different helioseismic data sets.
We used their comparison to estimate the $r$-dependence of the systematic
error in~$c_\odot$ in the core and in the convective
envelope (namely, a fractional systematic
error decreasing linearly from~0.0013 at $r = 0.05\;R_\odot$
to 0.0003 at $r = 0.2\;R_\odot$, constant from there to $r = 0.72;R_\odot$,
then increasing linearly to~0.00052 at $r = 0.94;R_\odot$).
For~$c_\odot$, this systematic error can be significantly larger than the
statistical errors quoted in the Table~2 of
\citet{BasuPB00},
and we combined the two in quadrature to get the
fractional error~$( \sigma_c / c )$
for the purpose of calculating weighted rms differences --- the
rms fractional difference in~$c$ is given by
$\left( \left\{ \sum \left[ ( \Delta c / c )
   / ( \sigma_c / c ) \right]^2 \right\}
 / \left\{ \sum \left[ 1 / ( \sigma_c / c ) \right]^2 \right\} \right)^{1/2}$.
For $(\gammaone)_\odot$ and~$\rho_\odot$, the systematic errors are comparable
to or smaller than the statistical ones, and the statistical errors
sufficed for calculating weighted rms differences.

\section{Results and Discussion} \label{sec:results}

\subsection{Sound Speed and Density Profiles} \label{ssec:profiles}

\begin{figure}[tp]
%
%
 \epsscale{0.67}
 \plotone{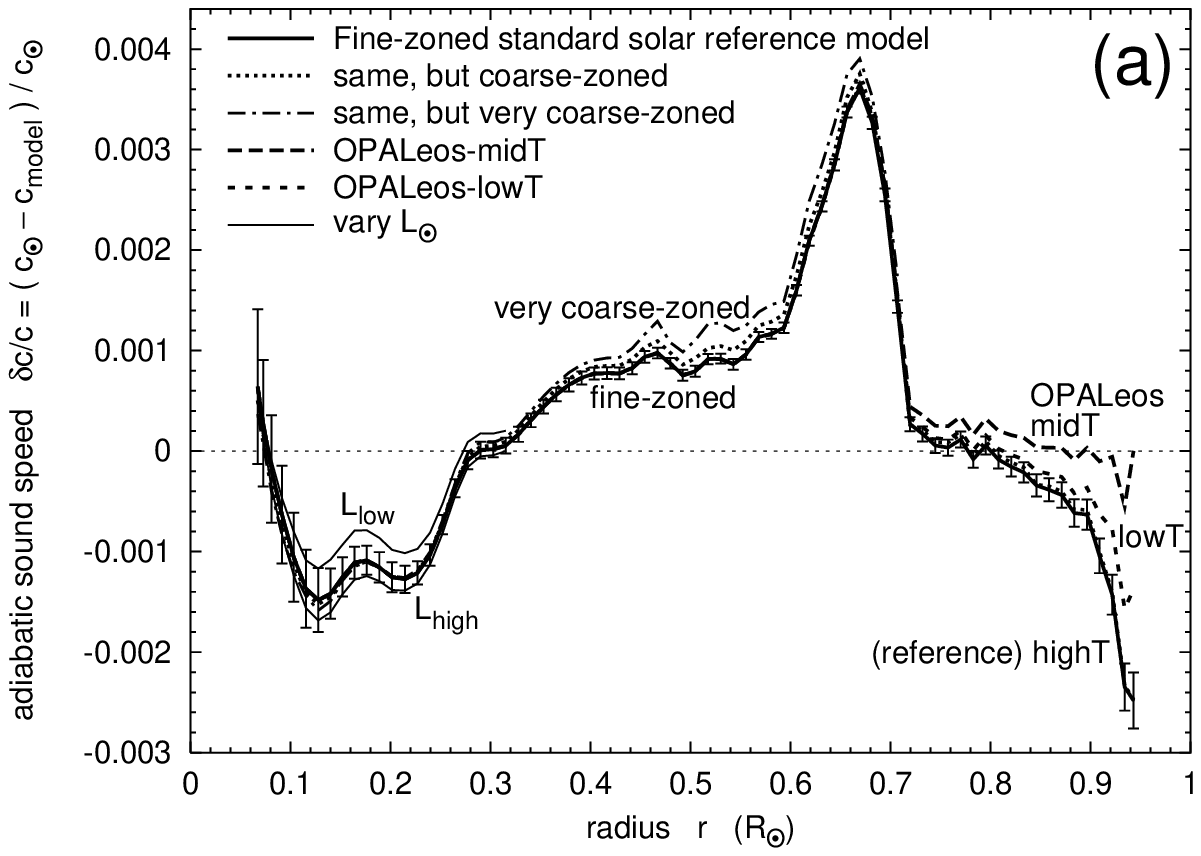}
 \plotone{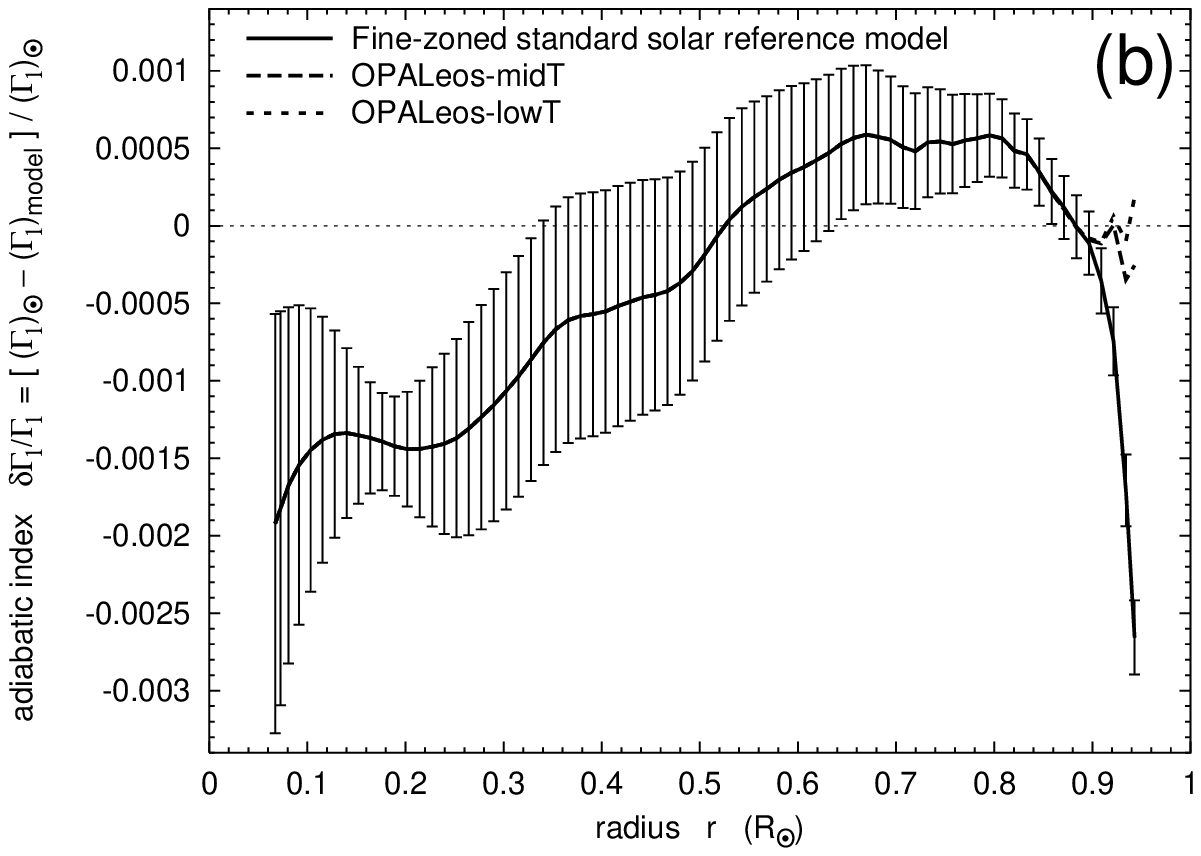}

\caption{Changing the zoning, the outer-envelope equation
of state, or the solar luminosity: the effects on
(a)~the adiabatic sound speed~$c$, and
(b)~the adiabatic index~$\gammaone$.
The reference standard solar model
({\it thick solid line\/}: errorbars give {\it statistical error only},
for the inferred helioseismic profile) switches
from the OPAL to the MHD equation of state for $\log\,\rho \lesssim -2$
(i.e., $r \gtrsim 0.94\;R_\odot$ or $\log\,T \lesssim 5.5$).
The ``OPALeos-midT'' ({\it short-dashed line\/}) and
``OPALeos-lowT'' ({\it dashed line\/}) models have the equation-of-state
switchover at $\log\,T \approx 4.0$ and $\log\,T \approx 3.75$,
respectively.  The {\it thin solid lines\/} show the effects of
using the maximum (``$L_{high}$'') and minimum (``$L_{low}$'') values
of~$L_\odot$; for clarity, these are shown only for the region
$r < 0.3\;R_\odot$ where there is a visible effect.}

 \label{fig:eos}

\end{figure}

\begin{figure}[t]
%
 \epsscale{0.67}
 \plotone{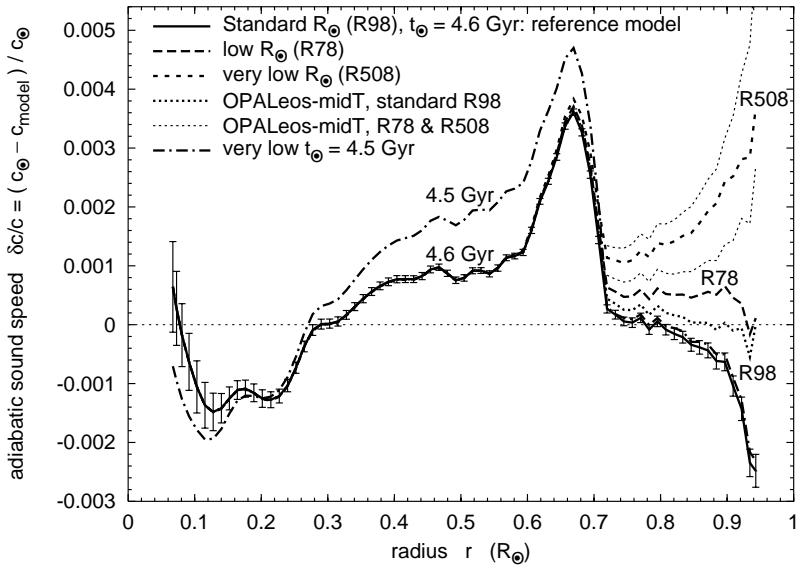}

\caption{Effect on the sound speed of changing
the solar radius or the solar age.
The reference standard solar model ({\it thick solid line\/})
has the standard solar radius ($R_\odot = 695.98\;$Mm: ``R98'')
and a (high) solar age ($t_\odot = 4.6$~Gyr).
The {\it dot-dashed line\/} shows the effect of using a {\it very\/}
low solar age $t_\odot = 4.5$~Gyr (the observational value of
$t_\odot = 4.57 \pm 0.01$~Gyr implies a total $3-\sigma$ age range
only half as large as the difference between these cases).
The {\it thick long-dashed line\/} shows the effect of using the low value
of $R_\odot = 695.78\;$Mm (``R78''), while the
{\it thick short-dashed line\/} shows the effect of using the very low value of
$R_\odot = 695.508\;$Mm (``R508'').
The {\it dotted lines\/} show the same radius comparisons for the
``OPALeos-midT'' case (in which the MHD equation of state is used
only for $\log\,T \lesssim 4.0$, rather than $\log\,T \lesssim 5.5$).
Note that variations in the solar radius would also affect the
{\it inferred helioseismic sound speed profile\/};
this is discussed in the text, but the effects are
{\it not\/} included in this figure.}

 \label{fig:trl}

\end{figure}

\begin{figure}[t]
%
 \epsscale{0.67}
 \plotone{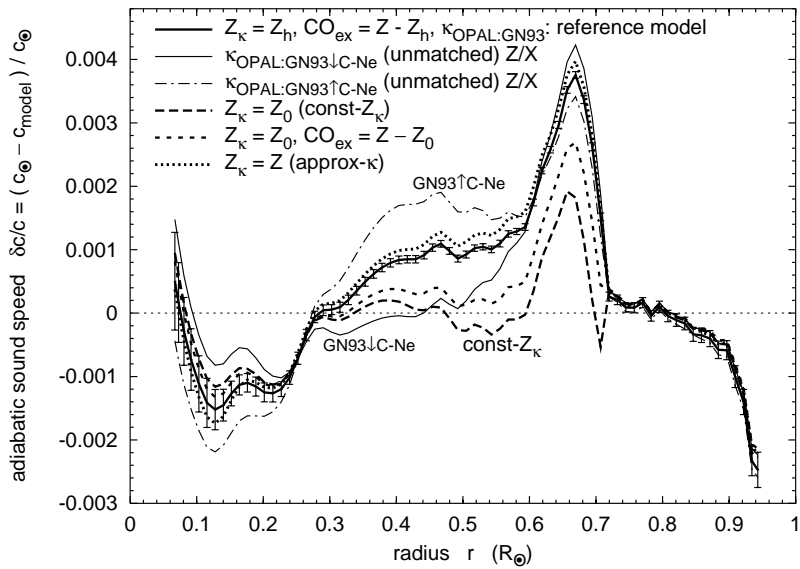}

\caption{Effect on the sound speed of opacity interpolation choices.
The {\it thin solid\/} and {\it dot-dashed lines\/} show the effects
of opacities where the C, N, O, and~Ne abundances were
respectively decreased (``$\kappa_{\rm OPAL:GN93\downarrow C-Ne}$'')
or increased (``$\kappa_{\rm OPAL:GN93\uparrow C-Ne}$'')
by their quoted errors of~15\%, but the
value of $Z/X$ was not re-adjusted to reflect these changes, so these
cases are not self-consistent (compare to the self-consistent cases
in Fig.~\ref{fig:ZXt} below).  Using opacity tables at a single constant
metallicity (``const-$Z_\kappa$:'' {\it long-dashed line\/}) leads
to large opacity errors and thus large sound speed errors, which cannot
be fixed by interpreting the metallicity error as ``excess-CO''
({\it short-dashed line\/}).  However, ignoring the effect on the
opacity of {\it relative\/} differences in the abundance profiles of the
individual metals results in only minor errors
(``approx-$\kappa$:'' {\it dotted line\/}).}

 \label{fig:kap}

\end{figure}

\begin{figure}[t]
%
 \epsscale{0.67}
 \plotone{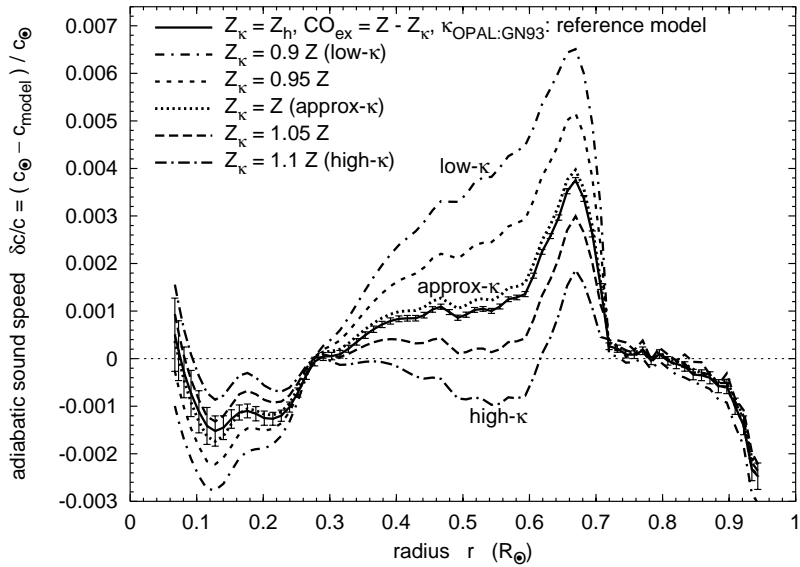}

\caption{Effect on the sound speed of opacity uncertainties.  The
{\it dot-dashed lines\/} (``low-$\kappa$'' and ``high-$\kappa$'')
show the effect of a $\sim 10$\% overall
change in the opacities relative to the reference standard solar model
({\it thick solid line\/}), while the {\it dashed lines\/} show the
effect of a $\sim 5$\% overall change in the opacities.}

 \label{fig:kapfz}

\end{figure}

\begin{figure}[t]
%
 \epsscale{0.67}
 \plotone{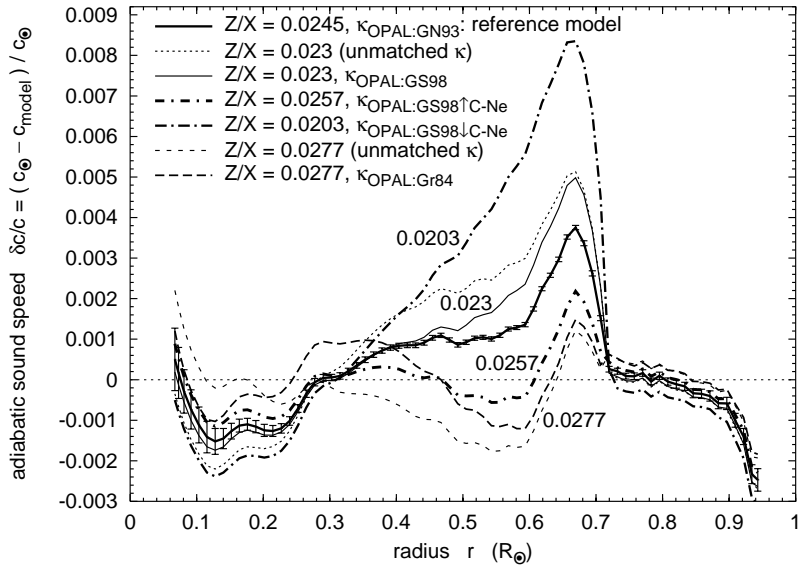}

\caption{Effect on the sound speed of uncertainties in the observed
solar surface composition.  Relative to the reference standard solar
model at $Z/X = 0.0245$ ({\it thick solid line\/}), switching to the
more recent $Z/X = 0.023$ value of
\citet{GreS98}
{\it without\/} including the corresponding changes in {\it relative\/}
metal abundances ({\it thin dotted line\/}) has a larger effect than the
full case with self-consistent abundances and opacities
(``$\kappa_{\rm OPAL:GS98}$:'' {\it thin solid line\/}); the same
is true when considering the older $Z/X = 0.0277$ cases of
\citet{Gre84}
({\it thin short-dashed line\/} vs.\ self-consistent
``$\kappa_{\rm OPAL:Gr84}$'' {\it thick long-dashed line}).
Nonetheless, a self-consistent test of the uncertainties in the
\citet{GreS98}
C, N, O, and~Ne abundances ({\it thick dot-dashed lines\/})
shows a large effect (compare to the {\it thin solid line\/}).}

 \label{fig:ZXt}

\end{figure}

\begin{figure}[t]
%
 \epsscale{0.67}
 \plotone{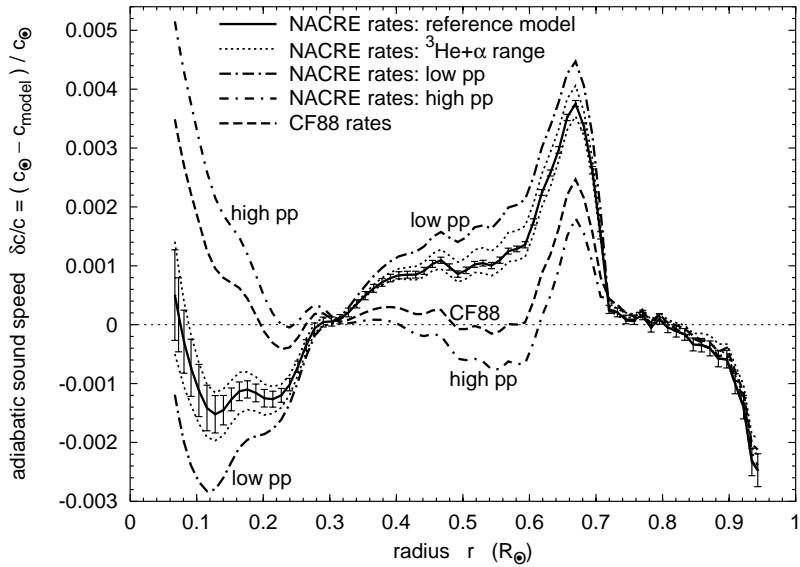}

\caption{Effect on the sound speed of uncertainties in nuclear rates.
The $\sim 5$\% uncertainty in the basic $pp$ rate
({\it dot-dashed lines\/}) has a much larger
effect than the 20\% uncertainty in the \iso3{He}${} + \alpha$ rate
({\it dotted lines\/}); other rate uncertainties have negligible
effects, and are not plotted.  The difference between the reference
standard solar model using the NACRE rates ({\it solid line\/}) and
the model using the
\citet{CF88}
rates (``CF88:'' {\it dashed line\/})
is largely due to the higher $pp$~rate adopted by the latter authors.}

 \label{fig:nuc}

\end{figure}

\begin{figure}[t]
%
 \epsscale{0.67}
 \plotone{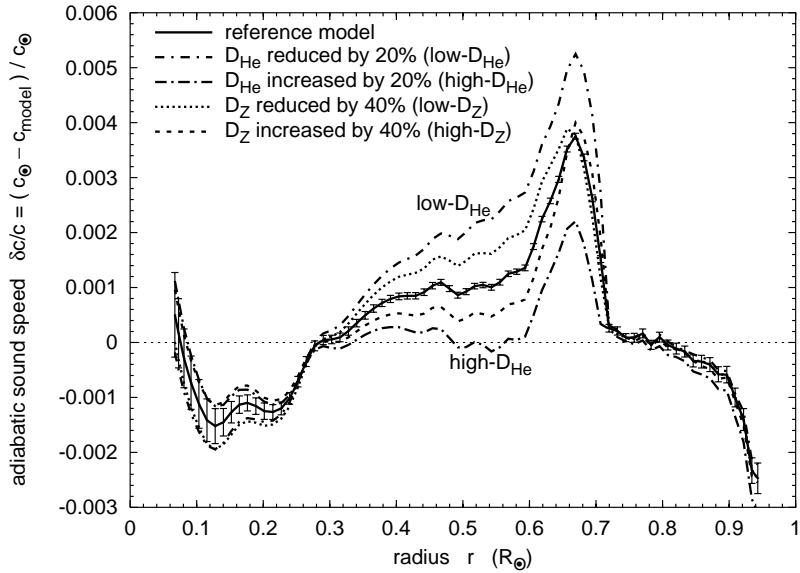}

\caption{Effect on the sound speed of uncertainties in diffusion constants.
The tested cases of a 20\% variation in the diffusion constant for helium
({\it dot-dashed lines\/}) may overestimate that uncertainty slightly;
on the other hand, the tested cases of a 40\% variation in the heavy element
diffusion constants ({\it dotted\/} and {\it short-dashed lines\/}) may
slightly underestimate the corresponding uncertainties --- the
reference standard solar model is shown by the {\it solid line}.}

 \label{fig:dif}

\end{figure}

We present in Figures \ref{fig:eos} through~\ref{fig:dif}
profiles of the adiabatic sound speed differences
$\delta c / c \equiv ( c_\odot - c_{model} ) / c_\odot$;
profiles of the density differences
$\delta \rho / \rho \equiv ( \rho_\odot - \rho_{model} ) / \rho_\odot$
are available online\footnote{
\url[http://www.krl.caltech.edu/~aib/papdat.html]{http://www.krl.caltech.edu/${}^{\sim}$aib/papdat.html}
}.
For our equation of state comparison, we also considered the equivalent
fractional difference in the adiabatic index~$\gammaone$.
Note that we use~``$\delta$'' to denote differences between the helioseismic
profile and one of our models, and ``$\Delta$'' to denote differences between
two of our models with different input parameters --- the ``$\delta$''~values
are the profiles plotted in our figures, while the ``$\Delta$''~values
refer to the difference between one plotted curve and another.

The theoretical sound speeds~$c_{model}$ and densities~$\rho_{model}$
are from our computed reference standard solar model and from our variant
standard solar models.  Our reference standard solar model used
current input parameters, as discussed in \S~\ref{sec:methods}; our
variant standard solar models comprised standard solar models with
one or more input parameters varied within the permitted range.
We present all our sound speed and density profiles in terms of
differences relative to the observed helioseismic reference profiles of
\citet{BasuPB00}.
This choice of presentation not only allows one to see the effects of
the uncertainties in the input parameters, but also shows which choice
of input parameters agrees best with the helioseismic observations.
Our ``OPALeos-lowT'' model (discussed in more detail below) is the one most
nearly comparable to the ``STD'' model of
\citet{BasuPB00},
and yields similar $\delta c/c$ and $\delta\gammaone/\gammaone$
curves, as may be seem by comparing their Figures~2 and~3 to our
Figure~\ref{fig:eos}.  Their models fit the solar sound speed profile
somewhat better than our models do, but not significantly so, considering
the size of the effects (discussed in detail further below)
that result from reasonable
variations in the input parameters of the solar model.  We made no attempt
to compare our solar models in the region outside $r = 0.943\;R_\odot$,
the last point on the helioseismic profiles of
\citet{BasuPB00}
--- the reason that this {\it is\/} their outermost point is that
significant systematic uncertainties arise in inversions near the
solar surface
\citep[see, e.g.,][]{DiMauro+02}.
However, we {\it have\/} compared the OPAL and MHD equation of state in
this region, as have some other investigators
\citep[see, e.g.,][]{GuzS97,Rich+98,BasuDN99,GongDN01,DiMauro+02};
this is discussed in detail further below.



\paragraph{Rotation effects:}
The prominent peak in $\delta c / c$
at $r \sim 0.7 \; R_\odot$ visible in
Figure~\ref{fig:eos}a is due to the
neglect of rotation-induced mixing just below the base of the solar
convective envelope, as has been shown by a number of investigators
who have included parameterized rotational mixing
\citep[see, e.g.,][]{Rich+96,BrunTZ99,BasuPB00,BahPB01,TurckC+01b}.
There are significant uncertainties in the physical processes that
lead to rotation-induced mixing.  However, all of
these investigators agree that rotational mixing is capable of smoothing
out the peak at $r \sim 0.7 \; R_\odot$, and that this has a relatively
small effect on the sound speed elsewhere in the Sun (a fractional
change of less than 0.001).  For example,
\citet{BahPB01}
found that including ``maximal'' rotational mixing spread out this peak
over the region $0.3\;R_\odot \lesssim r \lesssim 0.7\;R_\odot$,
eliminating the prominent peak but
worsening the agreement with the helioseismological sound speed
profile by about 0.001 in much of the solar interior
($0.3\;R_\odot \lesssim r \lesssim 0.6\;R_\odot$).
The ``minimal'' mixing model of
\citet{Rich+96}
yielded much the same result, as did similar models of other authors
\citep{Gabriel97,BrunTZ99,TurckC+01b}.
As far as the sound speed and density profiles in
the core and the convective envelope are concerned, rotational
mixing below the base of the convective envelope
should have no significantly effect, as shown by the above authors.

Since the prominent peak at $r \sim 0.7 \; R_\odot$ results from
the neglect of rotational mixing, we did not require agreement
in this region between profiles from our theoretical models and
profiles inferred from the helioseismic observations.  Nor did we
require agreement in the inner core region, since the present helioseismic 
observations still result in large uncertainties in the inferred
profiles there; for example, as shown by
\citet{BahPB01},
the use of a different helioseismic dataset could remove the
the sharp upturn in $\delta c / c$ for $r \lesssim 0.1 \; R_\odot$
in Figure~\ref{fig:eos}a (and even convert it into a downward trend).
On the other hand, we aimed for agreement in the regions
$0.1 \; R_\odot \lesssim r \lesssim 0.6 \; R_\odot$
and $0.72 \; R_\odot \lesssim r \lesssim 0.94 \; R_\odot$,
where disagreements
are due to imperfections in the input physics or uncertainties in
the observed solar parameters.
This is demonstrated by our variant models,
and by the variant models of other investigators
\citep[see, e.g.,][]{MorPB97,GuzS97,Rich+98,BasuPB00,GongDN01,NeuV+01a,%
NeuV+01b}.

\paragraph{Convergence accuracy effects:}
The accuracy with which the model is converged to the solar radius,
luminosity, and~$Z/X$ values can affect the sound speed profiles.
The extent to which this occurs depends on the accuracy of the
convergence, and on the sensitivity of the profiles to variations in
$R_\odot$, $L_\odot$, and~$Z/X$ (these are discussed in detail below).
Our convergence accuracy resulted in effects no larger than
a few parts in~$10^5$ on the sound speed in the solar interior
($r \lesssim 0.6\;R_\odot$).  In the convective envelope, where the
sound speed is quite sensitive to~$R_\odot$, the effect was typically
less than a part in~$10^4$,
but could be as high as a few parts in~$10^4$ for the few
models with the worst convergence in~$R_\odot$.

Note that the sound speed profile in most of the solar convective envelope
($0.72 \; R_\odot \lesssim r \lesssim 0.94 \; R_\odot$) is sensitive
mainly to the equation of state and to the solar radius, with other
uncertainties having only a minor effect there, as can be seen by
considering this subregion in Figures~\ref{fig:eos} through~\ref{fig:dif}.

\paragraph{Zoning effects:}
We investigated the effects of using two different zonings.  Our
coarse-zoned models had about 2000 spatial zones in the
model, and about 200 time steps in the evolution from the zero-age
main sequence to the present solar age (plus about 800 time steps on
the pre-main-sequence).  These values are comparable to those used
by most authors for solar models, although many authors use more complex
algorithms than we did
to compute changes between one timestep and the next (allowing
the use of fewer main-sequence timesteps at the cost of more CPU-time
per timestep),
and some authors ignore the pre-main-sequence evolution entirely (or use less
stringent accuracy conditions there).
Typically, we ran solar evolutionary sequences (iteratively improving the
input parameters $Z_0$, $Y_0$, and~$\alpha$)
until our coarse-zoned models were converged to match the
solar luminosity and radius to about a part in~$10^5$, and the
solar surface~$Z/X$ to a part in~$10^4$; a few cases where convergence
was slow were nearly 10 times worse.
Our fine-zoned models had $10\,000$ spatial zones and took 1500
main-sequence time steps (plus 6000 pre-main-sequence time steps)
--- a factor of 5 increase in both spatial and temporal precision ---
and were typically converged to better than a part in~$10^5$ for $R_\odot$
and~$L_\odot$, and a few parts in~$10^5$ for~$Z/X$.
We also tested some {\it very\/} coarse-zoned models, with
1000 spatial zones, 100 main-sequence time steps (plus 600
pre-main-sequence time steps), and convergence to the solar parameters
of a few parts in~$10^4$.
(A coarse-zoned converged solar model took a few hours
of CPU-time on a fairly high-performance ES40 computer, as compared to a few
days of CPU-time for a fine-zoned converged model; these times were
roughly tripled on a 450~Mhz Pentium~III PC\hbox{}.)

Figure~\ref{fig:eos}a shows that the fine zoning made only a very modest
improvement relative to the coarse-zoned case, less even than the
statistical errors in the sound speed and density profiles obtained from
helioseismic inversions.  Even the {\it very\/} coarse-zoned test case
did not do too badly: in the solar interior, it differs from the
fine-zoned case by no more than 0.0004 in the sound speed profile and
0.003 in the density profile (amounts comparable to the systematic
uncertainties in the helioseismic inversion) --- the rms
differences (over the entire Sun) are even smaller, namely
rms$\{\Delta c / c\} \approx 0.0003$ and
rms$\{\Delta \rho / \rho\} \approx 0.002$.  The coarse-zoned
model did about 3~times better still, with rms
differences relative to the
fine-zoned case of rms$\{\Delta c / c\} \approx 0.0001$ and
rms$\{\Delta \rho / \rho\} \approx 0.0008$.  Zoning changes had {\it no\/}
effect on the adiabatic index~$\gammaone$ (the coarse-zoned $\gammaone$
curves were not plotted in Fig.~\ref{fig:eos}b, since they would be precisely
superimposed on the fine-zoned curves).  Additional tests demonstrated
that changes in the coarseness of zoning always led to the {\it same\/}
negligibly small systematic shift in solar interior sound speed and
density values (although
inaccuracies in matching the observed solar surface parameters could
lead to slightly larger random variations in the convective envelope
region $r \gtrsim 0.7 \; R_\odot$).  We therefore felt justified in
running most of the models with our coarse zoning.  Note that
\citet{MorPB97},
with about 1000 spatial zones, 60 main-sequence time steps, and convergence
to present solar surface parameters of a part in~$10^4$ (similar to our
very-coarse-zoned case), claimed
a numerical internal accuracy of 0.0005 in the sound speed, similar to
what we found for our very-coarse-zoned case.

\paragraph{Equation-of-state effects:}
\citet{GongDN01} compared four current equations of state: their own
MHD equation of state
\citep{Dap+88},
the OPAL equation of state
\citep{RogSI96},
the CEFF equation of state
\citep{ChrD92},
and the SIREFF equation of state
\citet{GuzS97}.
In each case, they used the approximation of a 6-element composition mixture
(H,~He, C, N, O,~Ne) with the same composition as in the OPAL equation of
state of
\citet{RogSI96}.
For $3.7 \lesssim \log\,T \lesssim 6$, at corresponding solar densities
but {\it without\/} the usual ``$\tau$-correction'' that eliminates the
short-range divergence in the Debye-H\"uckel potential,
\citet{GongDN01}
find OPAL${} - {}$MHD differences in the equation of state of
$\Delta P / P \le 0.0005$, $\Delta \chi_\rho \le 0.005$,
$\Delta \chi_T \le 0.007$, and $\Delta \gammaone \le 0.004$,
with differences several times smaller at $6 \lesssim \log\,T \lesssim 7$,
and comparable differences between other pairs of equations of state ---
recall that $\chi_\rho \equiv ( \partial \ln P / \partial \ln \rho )_T$,
$\chi_T \equiv ( \partial \ln P / \partial \ln T )_\rho$, and the
adiabatic index is
$\gammaone \equiv ( \partial \ln P / \partial \ln \rho )_s$.
{\it With\/} the usual ``$\tau$-correction,'' but for a pure hydrogen-helium
mixture ($Z = 0$), they find much larger differences:
$\Delta P / P \lesssim 0.002$, $\Delta \chi_\rho \lesssim 0.02$,
$\Delta \chi_T \lesssim 0.03$, and $\Delta \gammaone \lesssim 0.008$.

\citet{GuzS97}
compared the effect in different solar {\it models},
finding roughly the same OPAL${} - {}$MHD difference as the last of the
above comparisons (namely, the case with the ``$\tau$-correction''),
except that they find a much larger difference in the pressure; in the
region $r \gtrsim 0.95 \; R_\odot$ (i.e., $\log\,T \lesssim 5.5$), they
find $\Delta P / P \le 0.014$, $\Delta C_p / C_p \le 0.036$, and
$\Delta \gammaone \lesssim 0.006$.  Note that
these differences include the effects of slightly different temperature,
density, and composition profiles in the different solar models (which may
either increase or decrease the differences in the thermodynamic quantities,
since the solar models are re-adjusted to reproduce the present solar
luminosity, radius, and surface composition).

We performed our own OPAL${} - {}$MHD comparison, at {\it fixed\/}
temperature, density, and composition grid-points in the equation of
state tables --- the same comparison as that performed by
\citet{GongDN01}.
Our results are consistent with those of
\citet{GuzS97};
for a typical solar $\{T,\rho\}$ profile in the outer envelope
region, we found $\Delta P / P \le 0.017$, $\Delta C_v / C_v \le 0.03$,
$\Delta \chi_\rho \le 0.015$,
$\Delta \chi_T \le 0.032$, and $\Delta \gammaone / \gammaone \le 0.006$.

The MHD equation of state is obtained in a fully self-consistent manner
from the free energy; inaccuracies can arise only from deficiencies in
the formulas used to obtain the free energy
\citep{Dap+88,GongDZ01}.
However, for the OPAL equation of state
\citep{RogSI96},
we found that there were significant inconsistencies when we compared their
tabulated values of $\gammaone$, $\gammatwo/(\gammatwo-1)$,
and $(\gammathree-1)$ to values calculated from their tabulated values of
$P$, $C_v$, $\chi_\rho$, and $\chi_T$.  In the solar core, these
inconsistencies are very small (a few parts in~$10^4$), but in the
outer envelope ($r \gtrsim 0.94 \; R_\odot$, or $\log\,T \lesssim 5.5$)
we found inconsistencies as large as~3\% at grid-points that would be used
in the OPAL interpolation formulae
when computing thermodynamic quantities (although the grid-points
{\it nearest\/} to the solar $\{T,\rho\}$ locus have inconsistencies
of less than~1\%).  Generally, the size of these inconsistencies varied
smoothly in the OPAL grid, but in at least a few positions a few grid
spacings away from the solar $\{T,\rho\}$ locus,
there were sharp ``spikes'' where one of the thermodynamic quantities
had a ``glitch'' (an error of several percent) at a just a couple of adjacent
density and/or temperature points.  In addition, for 4 of the 8~lowest OPAL
$T$-grid points (at $3.71 \lesssim \log\,T \lesssim 3.77$), there is a
``sawtooth'' error: at every second density value, the tabulated quantities
are shifted systematically relative to the values at neighboring densities
and temperatures.  These shifts can be as large as~1\% at low densities
for $C_v$ and~$\chi_T$, and are always of order~0.1\% for~$P$.

A new OPAL 2001 equation
of state has recently become available\footnote{
\url[ftp://www-phys.llnl.gov/pub/opal/eos2001/]{ftp://www-phys.llnl.gov/pub/opal/eos2001/}}
\citep[see also][]{Rogers00,Rogers01},
which includes relativistic electron effects and
extends to both lower temperatures and higher densities then the
original OPAL equation of state.
Preliminary tests indicate that this OPAL 2001 equation of state has
larger but smoother inconsistencies in its tabulated thermodynamic
quantities, with few or no glitches, except in the extended
high-density region (where there are some very large ones) and
at the 2~lowest $T$-grid points ($\log\,T < 3.35$).

\citet{GongDN01}
found that the changing from a 6-element composition mixture
to a 15-element mixture has an effect an order of magnitude smaller
than the OPAL${} - {}$MHD differences (we also performed such tests,
and came to the same conclusion).
They nonetheless recommended the use of at least a 10-element mixture for
the best accuracy in an equation of state.  They also noted a couple of minor
deficiencies in the MHD equation of state that ``moves it away from both
helioseismically determined values and OPAL''
\citep{GongDN01}.

\citet{MorPB97}
compared solar {\it models\/} with the OPAL equation of state to ones with
the CEFF equation of state, and found a small but non-negligible
effect: sound speed differences $\Delta c/c$ of slightly over 0.001
and density differences $\Delta \rho/\rho$ up to 0.01.
Note that they had set the value of~$Z_{eos}$
used in their equation of state to a fixed value of~0.019; however, since
the equation of state is only weakly sensitive to~$Z$, this should
have only a minor effect on their models, and should not affect their
comparison of the two equations of state.  In general, the temperature,
density, and composition profiles would {\it all\/} be slightly different
between solar models with different equations of state, since the input
parameters $Z_0$, $Y_0$, and~$\alpha$ are adjusted individually for each
solar model to obtain the best fit to the present solar luminosity, radius,
and surface composition.

\citet{GuzS97}
presented more extensive solar model comparisons,
comparing both their own SIREFF equation of state and
the MHD equation of state to the OPAL equation of state
(they too used a fixed~$Z_{eos}$, of~0.02, but again this should not affect
the comparison).  They likewise found an effect $\Delta c/c \le 0.001$ at
$r \lesssim 0.95\;R_\odot$ (with differences up to~0.004 near the surface).
They also presented differences between the values of the pressure~$P$,
specific heat at constant pressure~$C_p$, internal energy~$U$, and adiabatic
index~$\gammaone$.  As mentioned above, near the solar surface
($r > 0.95\;R_\odot$) differences between models with different
equations of state
were relatively large, of order~1\%.  However, for $r \lesssim 0.9\;R_\odot$
\citet{GuzS97}
reported OPAL${} - {}$MHD differences between their solar models of
$\Delta P / P \le 0.0015$,
$\Delta U / U \le 0.002$,
$\Delta \gammaone / \gammaone \le 0.0007$, and
$\Delta C_p / C_p \le 0.006$; the
OPAL${} - {}$SIREFF differences were slightly smaller
for~$C_p$, slightly larger for $P$ and~$U$, and much larger (a factor
of~$\sim 3$) for~$\gammaone$.  For $r \gtrsim 0.3\;R_\odot$, the SIREFF
value of~$\gammaone$ has several relatively large ``wiggles'' (fractional
variations $\sim \pm 0.002$) relative to either OPAL or MHD\hbox{}.
On the other hand, most of the difference in the core,
and perhaps some difference in the average trend further out, may be
due to the fact that SIREFF includes relativistic electron effects,
while MHD and OPAL do not --- although they {\it are\/} included in
the {\it new\/} OPAL 2001 equation of state
\citep{Rogers00,Rogers01},
and have recently been added to the MHD equation of state by
\citet{GongDZ01}.

\citet{EllK98}
estimated that inclusion of relativistic effects would reduce the MHD or OPAL
value of~$\gammaone$ by a fraction~0.002 at $r \approx 0.1\;R_\odot$, this
correction growing smaller with increasing~$r$, to reach 0.001 at
$r \approx 0.3\;R_\odot$ and zero near the solar surface.  They pointed out
that such a shift in~$\gammaone$ for models using the OPAL or MHD equation
of state would significantly improve the agreement in the solar interior
with the inferred helioseismic $\gammaone$ profile.  Certainly, if such
a correction were applied to our $\gammaone$~curve in Figure~\ref{fig:eos}b,
the model profile would agree with the helioseismic profile within the
statistical errors for all $r \lesssim 0.6\;R_\odot$ (recall that decreasing
a model quantity shifts the curve upwards in the figures).  Recently,
\citet{GongDZ01}
confirmed that adding relativistic electron effects to the MHD equation
of state yields a change in~$\gammaone$ very close to that estimated by
\citet{EllK98}.

The adiabatic sound speed is defined as $c = ( \gammaone P / \rho )^{1/2}$;
changes in the solar ratio of $P / \rho$ would result not only from changes
in the equation of state but also from readjustments of the solar structure
in response to these changes, so it is not obvious a priori what effect
relativistic corrections would have on the sound speed.
Consideration of the effect from~$\gammaone$ alone suggest that
relativistic corrections might reduce the
slope at $r \lesssim 0.5\;R_\odot$ in the $\delta c / c$ curve of
Figure~\ref{fig:eos}a.  The sound speed differences presented by
\citet{GuzS97}
for their OPAL${} - {}$SIREFF comparison suggest that this would in fact
be the case, and that a fractional decrease of order~0.001 in the 
sound speed $c_{model}$ near the Sun's center (i.e., an increase of~0.001
in $\delta c / c$ there) would result from relativistic corrections
to the OPAL equation of state.

\citet{Rich+98}
compared solar models with the OPAL and MHD equations of state,
looking at the value of~$\gammaone$ in the convective envelope
($0.72\;R_\odot \lesssim r \lesssim 0.98\;R_\odot$); they found
that the OPAL equation of state appeared to perform slightly better there.
This is not very surprising; what is perhaps more surprising is
how well the MHD equation of state does in the solar interior,
since it was only originally designed to be accurate for
$\rm \rho \lesssim 10^{-2}~g\;cm^{-3}$
\citetext{D.~Mihalas 1999, private communication;
\citealp[see also][]{Dap+88,GongDN01}}
--- note that this $\rho$~condition corresponds to
$r \gtrsim 0.94 \; R_\odot$ and $\log\,T \lesssim 5.5$ in the Sun.
We investigated the effect of
changing the equation of state only in this outer region where both are
expected to be valid.
Note that, while the OPAL {\it opacity\/}
tables are unreliable for $\log\,T \lesssim 4$ due to their neglect of
molecular opacities, the OPAL {\it equation of state\/} tables include
molecular hydrogen effects, and should be reasonably accurate down to
their lower tabulation limit of $\log\,T = 3.699$
\citep{RogSI96}.

\citet{BasuDN99}
considered the outer part of the solar convective envelope
($0.8\;R_\odot \lesssim r \lesssim 0.99\;R_\odot$); they used a helioseismic
inversion to study {\it intrinsic\/} $\gammaone$ differences relative to
the solar values, where $(\delta\gammaone/\gammaone)_{int}$ is only that
part of the difference that is ascribed
to the equation of state in the inversion.  They found that the OPAL
$\gammaone$ value was preferable to the MHD value for
$r \lesssim 0.97\;R_\odot$.  Both equations of state did relatively well in
the inner envelope, with $(\delta\gammaone/\gammaone)_{int} \lesssim 0.0005$;
but near the solar surface both yielded a relatively large (negative) peak
with $(\delta\gammaone/\gammaone)_{int} \sim 0.004$.  This peak was deeper
in for the MHD equation of state (at $r \sim 0.96\;R_\odot$, as compared to
$r \sim 0.975\;R_\odot$ for OPAL), and also wider
(reaching in to $r \sim 0.9\;R_\odot$, as opposed to $r \sim 0.95\;R_\odot$
for OPAL).
\citet{DiMauro+02}
used recent observations of high-degree modes to perform a similar
comparison, finding essentially the same result, though with a slightly
higher peak difference $(\delta\gammaone/\gammaone)_{int} \sim 0.0055$
near the surface.
Our Figure~\ref{fig:eos}b plots the {\it total\/} $\gammaone$ differences
rather than the intrinsic ones, but it is nonetheless likely that the sharp
downturn in our MHD $\delta\gammaone/\gammaone$ profile (solid curve)
at $0.9\;R_\odot \lesssim r \lesssim 0.94\;R_\odot$ corresponds
to the inner edge of this ``peak'' in the MHD disagreement.

Our reference standard solar model used the
OPAL equation of state in the interior regions,
switching over to the MHD equation of state in the outer envelope;
this switchover was performed gradually over the region
$-1.5 > \log\,\rho > -2$ (corresponding to $0.89\;R_\odot < r < 0.94\;R_\odot$
and $5.8 > \log\,T > 5.5$ in the present Sun).
We compared this reference standard model with two cases
where the switchover occurred even further out in the envelope:
a case ``OPALeos-midT'' where the switchover
occurred for $4.0 > \log\,T > 3.9$,
and a case ``OPALeos-lowT'' where the switchover occurred for
$3.75 > \log\,T > 3.7$.
In this latter ``OPALeos-lowT'' case, the MHD equation of state is used
only outside the Sun's photosphere, and thus has negligible effect
on the main sequence evolution.  On the other hand, any artifacts induced
by the switchover might be smaller in the former ``OPALeos-midT'' case, since
differences between the two equations of state are significantly
less near $\log\,T = 4$ than near $\log\,T = 3.7$ (although any such
artifacts should be small in any case, comparable to effects of the
inconsistencies in the OPAL equation of state,
as discussed in \S~\ref{sec:methods}).

Figure~\ref{fig:eos} shows that, as one would expect,
changing the equation of state in the
outer envelope alone has no effect on the interior, and in fact
there is only a
minor effect in that part of the convective envelope where the
equation of state remains unchanged.
(This also demonstrates that any artifacts from the equation of state
switchover do not affect the solar interior, and also have {\it at most\/}
a minor effect on the convective envelope.)
Since the effects were so small, we computed fine-zoned cases for this
equation-of-state test; these are the ones presented in
Fig.~\ref{fig:eos}.

Note that $\gammaone$ is affected significantly {\it only\/} by
variations in the equation of state.

\paragraph{Solar luminosity effects:}
\citet{BahPB01}
tested the effect of 2-$\sigma$ changes in the value of~$L_\odot$
(namely, $\pm 0.8$\%) on their solar models, finding only a minor effect
on neutrino fluxes and negligible effects on the other quantities they
considered.  For completeness, we made the same test with our own models,
confirming their results.  Figure~\ref{fig:eos}a shows the effects on the
sound speed of a solar luminosity 0.8\%~lower (``$L_{low}$'') and
0.8\%~higher (``$L_{high}$'') than the most recent value of
$3.842 \times 10^{33}\;$erg$\rm\;s^{-1}$
\citep{BahPB01,FroL98,Crom+96}
--- note that our reference standard solar model lies closer to~$L_{high}$
than to~$L_{low}$, as it uses a slightly higher~$L_\odot$ value than the most
recent estimate (see \S~\ref{sec:methods}).  To avoid confusion with other
curves, the ``$L_{high}$'' and ``$L_{low}$'' curves are shown only in the
region where they differ the most, namely, $r \lesssim 0.3\;R_\odot$;
even in this region, a shift of~0.8\% in~$L_\odot$ produces a fractional
change in the sound speed of less than 3~parts in~$10^4$, dropping to
1~part in~$10^4$ for $r > 0.3\;R_\odot$.



\paragraph{Solar radius effects:}
\citet{BasuPB00}
demonstrated that using a solar radius different from
the standard value of $R_\odot = 695.98\;$Mm
could have a small but not completely
insignificant effect on {\it both\/} the sound speed profile inferred from
helioseismic inversions and that computed in solar models.
They found that using the 0.03\% smaller solar radius value
$R_\odot = 695.78\;$Mm (case ``R78'') suggested by the $f$-mode study of
\citet{Antia98}
would reduce the inferred helioseismic sound speed profile throughout
the Sun by about the same small fraction, namely~0.0003 (nearly
independent of position in the Sun); similarly, using the 0.07\% smaller
value $R_\odot = 695.508\;$Mm
(case ``R508'') suggested by the solar-meridian transit study of
\citet{BroC98}
would reduce the inferred helioseismic sound speed profile by~0.0007.
On plots such as ours of fractional differences
$\delta c / c \equiv ( c_\odot - c_{model} ) / c_\odot$,
such a reduction in ``$c_\odot$'' would shift all the curves downwards
by the given amounts.  This shift has {\it not\/} been performed in
Figure~\ref{fig:trl} --- we only display our $c_{model}$ values relative to
the $c_\odot$ values of
\citet{BasuPB00},
so as to allow comparisons between different solar models
--- but this effect {\it has\/} been included in our quoted rms
values relative to the Sun ``rms$\{\delta c / c\}$''
(see next paragraph, and Table~\ref{tab:results}).
\citet{BasuPB00}
also calculated that a change in the solar radius would result in
a non-uniform shift in the inferred helioseismic $\rho_\odot$
and~$(\gammaone)_\odot$ profiles, by amounts
comparable to the statistical errors in
these quantities; although these shifts are barely significant statistically,
in contrast to the shift in~$c_\odot$, for completeness their effects have
been applied to our rms values calculated relative to the helioseismic
profiles for the ``R78'' and ``R508'' cases.

Figure~\ref{fig:trl} illustrates the effect on the sound speed of
changing the value of~$R_\odot$ from the standard ``R98'' case to the
smaller ``R78'' and ``R508'' cases.  Only in the convective envelope
is the sound speed significantly affected,
with the largest effect being near the solar surface.
In the ``peak'' region just below the convective envelope,
these ``R78'' and ``R508'' curves differ from the
reference standard solar model by no more than a few parts in~$10^4$, and
this difference drops to about a part in~$10^5$ for $r < 0.6\;R_\odot$.
In the convective envelope ($r \gtrsim 0.72\;R_\odot$),
the ``R78'' case is an improvement on the reference model
(reducing envelope-only rms$_{env}\{\delta c / c\}$ from 0.0007
to~0.0004 when one includes the effect of the shift in the inferred
helioseismic profiles), but the ``R508'' case is worse
(rms$_{env}\{\delta c / c\} = 0.0011$).
For the ``OPALeos-midT'' case, reducing the solar radius {\it always\/}
worsens agreement in the convective envelope (rms$_{env}\{\delta c / c\}$
of~0.0003 is
increased to 0.0009 or~0.0018 for ``R78'' or ``R508,'' respectively).
Figure~\ref{fig:trl} also shows that
the effect on the sound speed profile of changes in the solar
radius adds linearly to effects from changes in the envelope equation of
state.

The overall rms and the rms in the interior are not much affected by
variations in the solar radius,
as may be seen from Table~\ref{tab:results}.  However, if relativistic
corrections had been included in the equation of state, the $\delta c / c$
profile in Figure~\ref{fig:trl} would probably have been less negative
at $r \lesssim 0.3\;R_\odot$; an overall downward shift in the whole profile
(such as results from the overall shift in the inferred helioseismic sound
speed profile for smaller $R_\odot$ values) would then probably lead to some
improvement in rms$\{\delta c / c\}$.  However, the effect would still
be much smaller than some of the other effects discussed below.

\paragraph{Solar age effects:}
Figure~\ref{fig:trl} demonstrates that the uncertainty in the solar
age~$t_\odot$ has only a very minor effect on the solar sound speed
profile --- note that the shift illustrated here, from
$t_\odot = 4.6$~Gyr to 4.5~Gyr, is much larger than the observationally
allowed range of solar ages, i.e.,
4.55~Gyr${} \lesssim t_\odot \lesssim 4.59$~Gyr, as
discussed in~\S~\ref{sec:methods}.  (These ages are defined to include
the pre-main-sequence; main sequence ages can be obtained by subtracting
0.04~Gyr).  The maximum allowed shift of 0.02~Gyr relative to the ``best''
solar age of 4.57~Gyr would yield negligibly small effects, namely,
rms$\{\Delta c / c\} \approx 0.0001$ and
rms$\{\Delta \rho / \rho\} \approx 0.001$, with maximum changes less than
twice these values.  Our results agree both qualitatively and quantitatively
with the age sensitivity found in the recent work of
\citet{MorPB97}.

\paragraph{Low-temperature opacity effects:}
Uncertainties in the low-temperature molecular opacities would not be
expected to have much effect --- in a convective region, such as the solar
convective envelope, the
structure is almost independent of the local opacity.  As expected, using the
\citet{Shar92}
molecular opacities (``$\kappa_{\rm Sharp}$'') below $10^4$~K rather than the
\citet{AlexF94}
molecular opacities (``$\kappa_{\rm Alexander}$'') led to essentially
identical sound speed and
density profiles --- the ``$\kappa_{\rm Sharp}$'' case is thus not plotted in
Figure~\ref{fig:kap}.  The rms differences are negligible,
less than a part in~$10^4$ for the sound speed and less than a part
in~$10^3$ for the density.  Only pre-main-sequence lithium depletion
was significantly affected (see \S~\ref{ssec:lithium}).



\paragraph{Interior opacity effects:}
Recently,
\citet{NeuV+01b}
compared models using the 1995 OPAL opacities
\citep{IglR96,RogSI96}
with models using an updated version of the LEDCOP opacities from Los Alamos
\citep{Magee+95};
under solar
conditions, these two sets of opacities differ by up to 6\% (just below the
base of the convective envelope), although the authors indicate that about
half of this difference is due to interpolation errors (from different
temperature grids on which the opacities are tabulated).  They find
fractional sound speed differences up to $\Delta c / c \sim 0.003$
between solar models using these different opacity tables.
\citet{MorPB97}
have also demonstrated the serious impact of opacity changes on the
sound speed and density profiles of solar models; they compared
the 1995 OPAL opacities
with the less-precise 1992 OPAL opacities
\citep{RogI92}
(albeit with models that neglected diffusion), finding that the
improved opacities made an improvement of up to 0.005 in the sound speed
and up to 0.03 in the density (see their models S1 and~S2).
\citet{BasuPB00}
compared a model with the 1995 OPAL opacities and the OPAL equation
of state
\citep{RogSI96}
to a model with the 1992 OPAL opacities and the cruder Yale equation of
state
\citep{Gue+92}
with the Debye-H\"uckel correction
\citep{BahBS68}
(their models did include diffusion); they likewise
found an effect of up to 0.005 in the sound speed and up to 0.03 in
the density, from the combination of these two changes in the input.
We found that an even larger improvement of up to 0.007 in the sound speed
resulted from changing from the even older Los Alamos (LAOL) opacities
\citetext{Keady 1985, private communication}
to the 1995 OPAL opacities, with an rms improvement of~0.004 as shown
in Table~\ref{tab:results}
(there is also an improvement of up to 0.04 in the density, with an rms
improvement of~0.02) --- however, such a large opacity change as this
overestimates the uncertainty in the 1995 OPAL opacities
\citep{RogI98}.

As pointed out by
\citet{MorPB97},
neglecting the opacity changes that result from metallicity variations
in the Sun would lead to significant errors --- e.g., errors of
up to 0.0015 in the sound speed.
In our reference standard solar model, we did our best to account for these
temporal and spatial variations in the opacity due to these composition
changes from diffusion and nuclear burning.  As discussed
in~\S~\ref{sec:methods}, the metallicity
value~$Z_\kappa$ that we used for metallicity interpolation in
the OPAL opacity tables was scaled according to the changes in the
elements heavier than oxygen (``$Z_\kappa = Z_h$,'' where $Z_h$ is
proportional to the heavy element abundance).  Several different
methods were tested to account for the fact that changes in the
CNO-element abundances (particularly in the solar core) are far from
being proportional to changes in the heavy elements.  As discussed
in~\S~\ref{sec:methods}, all such methods that we tested gave essentially 
identical results --- even omitting the CNO-correction entirely had almost
no effect, and only a negligible improvement resulted from a full
``CNO-interpolation'' case that included the full opacity effects of
CNO abundance variations quite accurately (by interpolating among several
separately-computed OPAL opacity tables with different CNO abundances).
For our reference standard solar model, we simply assigned the sum of
the non-proportional changes in the CNO abundances (namely, $Z - Z_h$)
to the ``excess carbon and oxygen'' interpolation variable of the OPAL
opacity tables (i.e., CO$_{ex} = Z - Z_h = Z - Z_\kappa$).

An alternative approximation (``approx-$\kappa$'') is to set
$Z_\kappa = Z$ (i.e., to interpolate the OPAL opacity tables in the
local metallicity~$Z$, but ignore effects of variations in the
makeup of~$Z$).  As shown in Figure~\ref{fig:kap}, this approximation
yields results almost identical to those of our reference standard
solar model; rms differences are rms$\{\Delta c / c\} \approx 0.0002$
and rms$\{\Delta \rho / \rho\} \approx 0.001$, with maximum differences
not very much larger.
\citet{MorPB97}
compared two different ways of estimating the value of~$Z_\kappa$
(both being similar but not identical to our ``approx-$\kappa$''
case); they likewise found only very minor differences in the sound speed
between their two methods, but much larger effects on the density
(see their models D3 and~D12).

We also tested cases where OPAL opacities had been calculated for mixes
in which the abundances of C, N, O, and~Ne were either all
increased by their quoted
errors of~15\% (``$\kappa_{\rm OPAL:GN93\uparrow C-Ne}$'')
or decreased by this amount (``$\kappa_{\rm OPAL:GN93\downarrow C-Ne}$''),
relative to their abundances quoted by
\citet{GreN93}.
Figure~\ref{fig:kap} illustrates an effect of up to~0.001 in the sound
speed from such a change.  However, we used a value of $Z/X = 0.0245$
for all of the models in Figure~\ref{fig:kap}, which is not strictly
consistent with such large abundance changes: since C, N, O, and~Ne
comprise the major portion of~$Z$, a 15\% change in their abundances
should correspond to a change of~$\sim 12$\% in~$Z/X$ as well.  We
discuss such a self-consistent opacity-plus-composition-plus-$Z/X$
variations under $Z/X$ effects below.

Even when using the most up-to-date OPAL opacities, one can still get
significant errors if one neglects the effect on opacity of $Z$-changes
(primarily due to diffusion).  The simplest case is to set
$Z_\kappa = Z_0$ (``const-$Z_\kappa$''), where $Z_0$ is the
protosolar metallicity; in effect, such a case
uses only the OPAL opacity tables relevant to the protosolar metallicity
and ignores the effect on the opacity of
any subsequent changes in the metallicity.
Figure~\ref{fig:kap} demonstrates that this ``const-$Z_\kappa$''
case yields errors of up to~$\Delta c / c \sim 0.0015$
relative to the more accurate opacity interpolation of the reference
standard solar model, in agreement with the results of
\citet{MorPB97}
(compare their models D10 and~D12);
the rms errors were rms$\{\Delta c / c\} \approx 0.0010$ and
rms$\{\Delta \rho / \rho\} \approx 0.007$.
One might attempt to fix up this neglect of metallicity variation
by interpolation using the mildly CO-enhanced OPAL opacity tables,
i.e., retaining a constant $Z_\kappa = Z_0$ for opacity interpolation
purposes but setting CO$_{ex}$ to the difference between~$Z_\kappa$
and the true value of~$Z$ that results from diffusion and nuclear
burning, i.e., CO$_{ex} = Z - Z_\kappa = Z - Z_0$.
Figure~\ref{fig:kap} illustrates
that this approximation is a slight improvement over the
``const-$Z_\kappa$'' case but still not very satisfactory: it still
has rms errors of rms$\{\Delta c / c\} \approx 0.0006$ and
rms$\{\Delta \rho / \rho\} \approx 0.004$.

Interpolation errors can also arise from the finite grid spacing of the
opacity tables in $X$, $Z$, $T$, and~$\rho$.  In creating the opacity
interpolation routines, we tested the $X$- and $Z$-interpolation, finding
that these should result in only minor errors (a fraction of a percent)
in the opacity.  The work of
\citet{NeuV+01b}
suggests that $T$- and $\rho$-interpolation errors in the opacity
can be larger, as much as a few percent.

In addition to errors introduced by methods of interpolating in opacity
tables, one must consider the errors in the actual opacity values
contained in the tables.  Such opacity errors will in general be functions
of temperature and density; also, different elements will have different
errors.
\citet{RogI98}
estimate that there is a 4\% uncertainty in the 1995 OPAL opacities from
effects neglected in their calculations ---
\citet{NeuV+01b}
point out that the $\sim 3$\% intrinsic differences between the OPAL
and LEDCOP opacities are slightly less than this.
\citet{Turc+98}
showed that differences in the Rosseland mean opacities between 1992 and
1995 OPAL opacities do not exceed 7\% over the run of temperature and
density in the Sun's interior, and that these opacity differences
yielded sound speed differences of up to 0.002 in their solar models.
In addition, errors in the observed relative heavy element
abundances in the solar envelope will translate into opacity errors,
since different elements have somewhat different opacities.
\citet{RogI98}
estimate that such abundance uncertainties correspond
to opacity uncertainties of order~5\% at temperatures where ionization
effects of the relevant elements yield a large contribution to the
opacity (e.g., near $2 \times 10^6$~K for oxygen or neon).
\citet{Turc+98}
found that taking into account the
opacity effects due to changes in the relative
abundances of all the individual elements in~$Z$
led to opacities that differed by up to~2\% from the opacities
tabulated for the standard scaled-solar metallicity, yielding sound
speed differences of up to~0.001 and density differences up to~0.005 --- these
are several times larger than the effects discussed above that we found
when testing effects of variations of C, N, and~O relative to the
heavier elements.



We did not attempt detailed element-by-element variations of the OPAL
opacities in our models; nor did we test the effect of opacity variations
in limited density or temperature ranges.
Instead, we obtained a rough estimate of the maximum possible effects of
uncertainties in heavy-element opacities by making an overall shift
in the metallicity value~$Z_\kappa$ used for interpolation in the OPAL
opacity tables.  Figure~\ref{fig:kapfz} illustrates the cases
$Z_\kappa = 0.9 \, Z$ (``low-$\kappa$''), $Z_\kappa = 0.95 \, Z$,
$Z_\kappa = Z$ (``approx-$\kappa$''), $Z_\kappa = 1.05 \, Z$,
and $Z_\kappa = 1.1 \, Z$ (``high-$\kappa$'').  The ``low-$\kappa$''
and ``high-$\kappa$'' cases correspond to
an average shift in the opacities of order~10\% over the solar interior
relative to the ``approx-$\kappa$'' case
($2-5$\% for $r \lesssim 0.4\;R_\odot$, $\sim 10$\% for
$0.4\;R_\odot < r < 0.7\;R_\odot$, $\sim 15$\% for
$0.7\;R_\odot < r < 0.92\;R_\odot$, and $\sim 5$\% for
$r > 0.92\;R_\odot$).  As may be seen from Figure~\ref{fig:kapfz}, such
an opacity shift of order~10\% yields sound speed changes of up to
$\Delta c / c \sim 0.003$, with rms$\{\Delta c / c\} \approx 0.0016$
and rms$\{\Delta \rho / \rho\} \approx 0.014$.  However, such a large
opacity shift almost certainly overestimates the effects of opacity
uncertainties.
\citet{RogI98}
estimate that there is a 4\% uncertainty in the 1995 OPAL opacities from
effects neglected in their calculations, and it would be surprising if
these yielded a uniform shift in the opacity throughout the Sun.  Thus
a better estimate of the effects of opacity uncertainties would be
$\Delta c / c \sim 0.001$ and rms$\{\Delta \rho / \rho\} \sim 0.005$.



\paragraph{Solar abundance (Z/X) effects:}
Our reference standard solar model used the observational value
of $Z/X = 0.0245$ from
\citet{GreN93},
since their mixture was the one for which the standard OPAL opacity
tables (``$\kappa_{\rm OPAL:GN93}$'') were available.
The close-dashed curve in Figure~\ref{fig:ZXt} demonstrates the effects of
using the
\citet{GreN93}
relative metal abundances with the corresponding ``$\kappa_{\rm OPAL:GN93}$''
OPAL opacities, but using a 13\% higher value of $Z/X = 0.0277$, the
older value that had been recommended by
\cite{Gre84};
the maximum sound speed difference relative to the reference standard solar
model is $\Delta c / c \sim 0.0030$, with
rms$\{\Delta c / c\} \approx 0.0018$
and rms$\{\Delta \rho / \rho\} \approx 0.017$.
The dotted curve in Figure~\ref{fig:ZXt} illustrates a similar case with
a 6\% lower value of $Z/X = 0.023$, as recommended by the more recent work of
\citet{GreS98};
it has maximum $\Delta c / c \sim 0.0016$, with
rms$\{\Delta c / c\} \approx 0.0010$
and rms$\{\Delta \rho / \rho\} \approx 0.009$.
(Note that most of the above effect comes from the different opacity that
results from the changed solar $Z$ value, as may be seen by comparing with
the ``high-$\kappa$'' and ``low-$\kappa$'' curves in Fig.~\ref{fig:kap}.)
However, these comparisons are not strictly self-consistent, since it is
the changes in the individual elemental abundances of the metals that
add up to yield the changed $Z/X$ ratio.  Using the old abundance pattern of
\citet{Gre84}
and newly-computed OPAL opacities appropriate to it
(``$\kappa_{\rm OPAL:Gr84}$'') leads to the wide-dashed $Z/X = 0.0277$
curve in Figure~\ref{fig:ZXt}, with a slightly smaller
maximum sound speed difference (of~0.0025), and
rms$\{\Delta c / c\} \approx 0.0014$
and rms$\{\Delta \rho / \rho\} \approx 0.011$.  Similarly,
using the
\citet{GreS98}
abundance pattern and appropriate OPAL opacities
(``$\kappa_{\rm OPAL:GS98}$'') leads to the thin solid $Z/X = 0.023$ curve
in Figure~\ref{fig:ZXt}, reducing the maximum sound speed difference to
0.0012, with similarly reduced
rms$\{\Delta c / c\} \approx 0.0006$
and rms$\{\Delta \rho / \rho\} \approx 0.004$.

\citet{NeuV+01a}
performed a comparison identical to this last case, finding essentially
the same effect on the sound speed, both qualitatively and quantitatively
(maximum effect $\sim 0.0018$).
\citet{MorPB97}
considered the effect of a 6\% increase in~$Z/X$, finding a maximum
difference of 0.0007 in their sound speed and 0.003 in their density;
this would imply a significantly lower sensitivity to~$Z/X$ than we
found.  This is probably due to the fact that the models in which they
tested $Z/X$ variations did not consider the effect on the opacities
of the temporal and spatial variations in the heavy element abundances
that arise from diffusion, but merely used opacities appropriate to a
constant metallicity equal to the protosolar value ($Z_\kappa = Z_0$,
as in our ``const-$Z_\kappa$'' case discussed above).

Strictly, the uncertainty resulting from observational solar abundance
errors can be estimated by varying the solar abundance values of
\citet{GreS98}
within their quoted uncertainties, obtaining OPAL opacities
with these revised compositions, calculating the resulting $Z/X$ values,
and running solar models with these self-consistent sets of input
values.  We have done this for two cases.  Rather than performing large
numbers of random variations of the abundances, we tested a case which
should give something close to the maximum effect.  The elements
C, N, O, and~Ne not only comprise the major part of the metallicity but
also have relatively large errors of~$\sim 15$\%, and unlike other
elements with large errors one cannot get a ``better'' value by using
the meteoritic abundance instead.  We therefore considered cases where
C, N, O, and~Ne were either all increased by~15\% ($Z/X = 0.0257$,
``$\kappa_{\rm OPAL:GS98\uparrow C-Ne}$'') or all decreased by~15\%
($Z/X = 0.0203$, ``$\kappa_{\rm OPAL:GS98\downarrow C-Ne}$'') --- i.e.,
these self-consistent abundance variations correspond to 12\%
variations in~$Z/X$.  As
illustrated in Figure~\ref{fig:ZXt}, these cases lead to variations
in the solar sound speed of up to~0.003 relative to the
$Z/X = 0.023$ ``$\kappa_{\rm OPAL:GS98}$'' case,
with rms$\{\Delta c / c\} \approx 0.0017$
and rms$\{\Delta \rho / \rho\} \approx 0.012$.
Of all the ``input'' uncertainties that we
considered, these uncertainties in the solar abundances have the
largest impact.



\paragraph{Nuclear rate effects:}
Figure~\ref{fig:nuc} demonstrates that the
uncertainty in the \iso1H$(p,\nu e^{+})$\iso2H
nuclear burning rate (the $pp$~rate) has a significant impact
on the solar sound speed.
Our reference standard solar model used the recommended nuclear
rates from the NACRE compilation
\citep{Ang+99}.
These authors also supply ``high'' and ``low'' cases to indicate the
allowed uncertainty range of each nuclear rate; in the case of the
$pp$~reaction, the high case is 8\% above the recommended rate
and the low case is 3\% below it.  We have tested the
effects of nuclear rate uncertainties by computing variant standard
solar models using high and low NACRE rate values.  Figure~\ref{fig:nuc}
demonstrates that a high $pp$~rate is preferable, if all other
parameters are kept constant: the high $pp$~rate gives good agreement
with the helioseismic reference profiles, except in the Sun's central
regions where the helioseismic observations are the poorest.
Our models indicate that a change of 5\% in the
$pp$~rate yields changes of up to 0.003 in the sound speed,
(0.0014 in the regions accurately probed by helioseismology, outside
the core); the rms changes in such a case would be
rms$\{\Delta c / c\} \approx 0.0009$
and rms$\{\Delta \rho / \rho\} \approx 0.018$.
\citet{AntC99}
also tested the effects of changes in the $pp$~rate on the
helioseismic profiles, concluding that a relatively high $pp$~rate is
preferred, consistent with our results discussed above.

Figure~\ref{fig:nuc} also demonstrates
that the uncertainty of $\pm 20$\% in the
\iso3{He}$(\alpha,\gamma)$\iso7{Be} reaction leads to only a minor
effect: a maximum sound speed change of 0.001 (or 0.0003 outside the core),
with 
rms$\{\Delta c / c\} \approx 0.0002$
and rms$\{\Delta \rho / \rho\} \approx 0.004$.
\citet{BasuPB00}
considered the effect of setting the \iso3{He}$(\alpha,\gamma)$\iso7{Be}
rate to zero; they found large effects from such an unphysically
extreme change.  Setting the rate to zero is equivalent to a
100\% change, 5~times as large as the 20\% change that we considered; thus
it is consistent that their published effect is about 5~times as
large as ours.

We also tested the effects of the $\pm 6$\% uncertainty in the
\iso3{He}$(\iso3{He},2p)$\iso4{He} reaction and of the
$\pm 30$\% uncertainty in the \iso{14}N$(p,\gamma)$\iso{15}O reaction
(which determines the CNO-cycle rate).  Such changes in these rates led to
negligible effects on the sound speed and density profiles; we have not
plotted these profiles in Figure~\ref{fig:nuc}, since they would be
essentially superimposed on that of the reference
standard solar model.

It is not surprising that the uncertainty in the $pp$~rate has the
largest effect on the sound speed and density profiles, since it is
the basic rate that determines the overall \hbox{p-p} chain burning
rate.

We also computed a model using the previous standard set of nuclear
rates, namely, the Kellogg nuclear rate compilation of
\citet{CF88}.
The resulting sound speed and density profiles are shown in
Figure~\ref{fig:nuc}.  With us still in Kellogg, carrying out this
work in an office directly below his long-time office,
it is especially gratifying for us to see that Willy Fowler's
last published $pp$~rate yields such good agreement with the current
helioseismic reference profiles (the
largest differences being near the center, where the observations
are least accurate).

\paragraph{Electron screening effects in nuclear rates:}
\citet{GruB98}
performed a careful quantum mechanical computation of the effects of
electron screening on nuclear reactions in the Sun, and found that the
\citet{Sal55}
formula leads to only a very slight overestimate of the nuclear rates:
0.5\%~for the $p+p$ reaction, 1.7\%~for the $\iso3{He}+\iso4{He}$
reaction, 1.5\%~for the $p+\iso7{Be}$ reaction, and 0.8\% for the
$p+\iso{14}N$ reaction.  Since these corrections are an order of magnitude
smaller than the uncertainties given in the corresponding NACRE nuclear rates
\citep{Ang+99},
there was no point in considering separately the uncertainty in the
screening corrections.  Note that if one used the intermediate screening
formulae (including the partial degeneracy correction) of
\citet{Grab+73},
one would overestimate the nuclear reaction rates in the Sun by a
further 1~to~3\% relative to the weak screening formula ---
\citet{GruB98}
quoted a much larger effect in the opposite direction, but this was for a
version of the
\citet{Grab+73}
formulae that assumed completely degenerate electrons.

\citet{ShavS96}
suggested that electron cloud-cloud interactions would increase the
electron screening factor in the exponent of nuclear rates by a factor
of~2/3; this was based on a ``fundamental misconception concerning
the dynamics of the interaction''
\citep{BrugG97}.
\citet{CarSK88}
suggested that ``dynamic screening'' should reduce the
\citet{Sal55}
screening factor, since half of the screening effect comes from ions,
which should
not be able to adjust to the rapid motion of colliding/fusing nuclei.
However,
\citet{Gru98}
gave a general argument showing that in an equilibrium plasma there should
be no such reduction, and
\citet{BrownS97}
showed explicitly that such a modification of the Salpeter screening factor is
exactly cancelled when one considers processes whereby Coulomb interactions
with the colliding particles cause plasma excitations and de-excitations.
\citet{Bah+02}
point out that the more recent ``dynamic screening'' calculations of
\citet{ShavS00,ShavS01}
seem likewise to be based on misconceptions.

\citet{Bah+02}
showed several ways of deriving the fact that the
\citet{Sal55}
electron screening formula gives the correct leading term under weak
screening conditions, and used the rigorous formulation of
\citet{BrownS97}
to show that, in the Sun, corrections to this leading term are small
(of order~1\%).  They also pointed out that Tsytovich's alternative
``anti-screening'' formula
\citep{Tsyt00,TsytB00},
which yields a reduction in nuclear rates rather than an increase, yields
unphysical results in two different limits.  Nonetheless, some authors
\citep{FioRV01,WeisFT01}
have looked into the effects of using Tsytovich's alternative
``anti-screening'' formula, and
have shown that it yields solar models that are not
compatible with the helioseismic sound speed profile.

\paragraph{Non-Maxwellian ion velocity distributions:}
\citet{Cora+99}
suggest that the high-energy tail of the Maxwellian ion velocity distribution
may be slightly depleted in the Sun; to first order, they express this
modified distribution by $f(E) \sim (kT)^{-3/2} e^{-E/kT-\delta(E/kT)^2}$,
where they estimate that $\delta$ should be of the order~0.01.  Such a large
distortion of the high-energy tail would reduce the $p+p$ reaction rate by
more than~20\% (and other $pp$-chain reaction rates by an order of magnitude);
as may be seen by considering Figure~\ref{fig:nuc}, such a large reduction
in the $p+p$ reaction rate would not be compatible with the helioseismic
sound speed profile.  In addition,
\citet{Bah+02}
have criticized the above estimate of the magnitude~$\delta$ of the effect,
claiming that it should be negligibly small.  Nonetheless,
\citet{TurckC+01b}
have considered the effect of such a Maxwellian distortion, with a smaller
value of $\delta = 0.002$ --- this reduces the $p+p$ reaction rate by
only~5\%, yielding a solar model whose sound speed differs by
$\delta c/c \lesssim 0.005$ from the helioseismic profile (i.e., disfavored
by helioseismology, but perhaps not ruled out entirely).



\paragraph{Diffusion effects:}
There are uncertainties in the diffusion coefficients;
\citet{Prof94}
estimates a 15\% uncertainty in the diffusion constant of helium
relative to hydrogen, and a $\sim 50$\% uncertainty in the diffusion
constant for oxygen relative to hydrogen.
Figure~\ref{fig:dif} shows that an increase or decrease of 20\% in the
helium diffusion constants has only a modest effect: a maximum change
of 0.001 in the sound speed, with
rms$\{\Delta c / c\} \approx 0.0008$
and rms$\{\Delta \rho / \rho\} \approx 0.007$.
The effect of increasing or decreasing the heavy element diffusion constants
by~40\% has a slightly smaller effect: a maximum change of 0.0006
in the sound speed, with rms$\{\Delta c / c\} \approx 0.0004$
and rms$\{\Delta \rho / \rho\} \approx 0.004$.
Note that including the effects of
different diffusion rates for different heavy elements would have
an effect less than half as large as this, as shown by the results of
\citet{TurcC98}
and
\citet{Turc+98}.

\subsection{Solar Convective Envelope Depth} \label{ssec:rce}

One of the key results of helioseismic observations is a highly precise
value for the position~$R_{ce}$ of the base of the solar surface
convective region:
\citet{BasuA97}
report a value of $R_{ce} = 0.713 \pm 0.001 \; R_\odot$.
Our reference standard solar model is in agreement with this value,
having $R_{ce} = 0.7135 \; R_\odot$ (see Table~\ref{tab:results}),
with values negligibly different ($R_{ce} = 0.7133$ or $0.7134 \; R_\odot$)
if we used the OPAL
instead of the MHD equation of state at $\log\,\rho \lesssim -2$.
The new
\citet{GreS98}
solar abundance observations (implying $Z/X = 0.023$) yield a
barely-consistent value of $R_{ce} = 0.7157 \; R_\odot$; the
maximum allowed abundance variations about this value
(i.e., low and high $Z/X$ values of 0.0203 and 0.0277, respectively) are
inconsistent with the observed value (yielding $R_{ce} = 0.7209$
and~$0.7098 \; R_\odot$, respectively).  The
solar age uncertainty of $\pm 0.02$~Gyr
does not make a significant difference in~$R_{ce}$ (see
Table~\ref{tab:results}).
As far as the uncertainties in the nuclear reaction rates are concerned,
only the $pp$~rate has a significant effect on~$R_{ce}$,
of $\pm 0.002 \; R_\odot$.  Uncertainties in molecular opacities do not have
a significant effect on~$R_{ce}$, but changing from the old 1985 LAOL
opacities to the 1995 OPAL opacities does yield a large improvement
(of~$0.006 \; R_\odot$); the remaining uncertainties in the 1995 OPAL
opacities might thus be expected to have a small but possibly significant
effect on~$R_{ce}$.  The uncertainty in the diffusion constant
for helium does have a significant influence ($\pm 0.003 \; R_\odot$), but
uncertainties in the diffusion constants for the heavy elements do
not (effects${} \lesssim 0.001 \; R_\odot$).

Note that the cases favored by the sound speed profiles at the one- to
two-sigma significance level (high opacities, high~$Z/X$, or high $pp$~rate)
are disfavored by the observed~$R_{ce}$ value at about the same
significance level.

\subsection{Solar Helium Abundance} \label{ssec:helium}

Another key result of helioseismic observations is a fairly precise value
for the present solar envelope helium mass fraction~$Y_e$ (this value
is lower than the Sun's initial helium abundance, due to diffusion).
Inferring the solar
helium abundance requires the use of a (theoretical) equation of state, as
well as helioseismic frequency observations of modes that probe the solar
convective region, particularly the \ion{He}{2} ionization zone
\citep{Rich+98}.

Table~\ref{tab:Yenv} quotes helioseismic $Y_e$~values from eight
recent papers --- note that most authors calculate two separate values,
obtained using the OPAL and MHD equations of state, respectively.
Clearly, the systematic errors are much larger than the quoted internal
errors from the helioseismic inversions, as shown by the relatively large
differences between values obtained with these two different
equations of state, and by the differences between determinations
by different investigators.
The OPAL equation of state is expected to be more accurate than the MHD
equation of state over the bulk of the convective envelope, since the
MHD equation of state was
designed for use at $\log\,\rho \lesssim -2$ (which occurs in the Sun
at $r \gtrsim 0.942 \; R_\odot$); however, the \ion{He}{2} ionization
zone occurs further out ($0.975\;R_\odot \lesssim r \lesssim 0.985\;R_\odot$)
in a region the MHD equation of state was
specifically designed for, and where it may actually be more accurate
than the OPAL equation of state
\citep[see, e.g.,][]{Rich+98,BasuDN99,DiMauro+02}.
Possible reasons for the relatively large scatter among $Y_e$
values of different investigators (even when using the same equation
of state) are discussed by
\citet{DiMauro+02};
they point out that observational frequencies for high-degree oscillation
modes may suffer from significant systematic errors, and that the inversion
formulae used to obtain~$Y_e$ ignore several physical effects that may
significantly affect the frequencies in this region of the Sun, such as
nonadiabaticity, effects of mode excitations, or flows resulting from
convective turbulence --- certainly there is evidence that turbulent motions
affect the $f$-modes
\citep{Gough93,ChitreCT98,MuraDK98,MedrMR99,DiMauro+02}.

One may perhaps ignore the extremely low value of $Y_e \approx 0.226$ of
\citet{ShiHT99},
since it was obtained not from a direct inversion but by fitting to a
sound speed profile (that had in turn been obtained via a helioseismic
inversion); relatively poor accuracy for this method is also suggested
by the fact that the simultaneous determination by these authors
of the depth of the solar convective envelope yielded
$R_{ce} \approx 0.718 \; R_\odot$, quite far from
helioseismic value of $R_{ce} = 0.713 \pm 0.001 \; R_\odot$
\citep{BasuA97}.
The remaining $Y_e$~values from
Table~\ref{tab:Yenv} have a mean of~0.245,
a median value of~0.248, and a total range of $0.23 \le Y_e \le 0.254$,
with most of the values lying in the range $0.24 \lesssim Y_e \lesssim 0.25$.
These results can be conveniently summarized as an
``observed'' value of $Y_e = 0.245 \pm 0.005$
(where most of the uncertainty is due to systematic effects).

Our theoretical reference standard solar model is in excellent agreement
with this, having $Y_e = 0.2424$ independent of whether we used the OPAL
or the MHD equation of state at $\log\,\rho \lesssim -2$.  Low and high
$Z/X$ values (0.0203 and 0.0277, respectively)
yield $Y_e$ values of~0.2396 and~0.2510, respectively --- still
quite acceptable.  Similarly acceptable changes of $\pm 0.005$ in~$Y_e$
result from a 4\% uncertainty in the opacities,
a 15\% uncertainty in helium diffusion constants, or a 50\% uncertainty
in heavy-element diffusion constants.
Uncertainties in the solar age, luminosity, and radius and in the
nuclear rates have only a negligible effect
on~$Y_e$ (see Table~\ref{tab:results}).

\subsection{Solar Lithium Abundance} \label{ssec:lithium}

The present observed solar surface lithium abundance is
$\log\,\varepsilon(\iso7{Li}) = 1.10 \pm 0.10$
as compared to the initial value of
$\log\,\varepsilon(\iso7{Li}) = 3.31 \pm 0.04$
obtained from meteorites
\citep{GreS98},
where $\log\,\varepsilon(\iso7{Li}) = \log ( N_{\rm Li} / N_{\rm H} ) + 12$
for number densities $N_{\rm Li}$ and $N_{\rm H}$ of lithium and hydrogen,
respectively.
The solar surface lithium depletion factor~$f_{\rm Li}$,
relative to its initial value, is thus observed to
be $f_{\rm Li} = 160 \pm 40$.  Solar surface lithium can be
depleted due to three causes: (1)~lithium burning during the
pre-main-sequence evolution, when the surface convection still reaches
deeply into the interior; (2)~rotationally induced mixing on the main
sequence, which transports lithium down from the convective envelope
to regions hot enough for lithium burning; (3)~mass loss on the main
sequence, which can cause the convective envelope to move inwards and
engulf lithium-depleted regions.  In this paper, we only consider the
first of these, namely, the pre-main-sequence lithium destruction;
rotational mixing is beyond the scope of this paper, and main sequence
mass loss is discussed in the companion ``Our Sun~V'' paper
\citep{SB02}.  Our reference standard solar model had a
pre-main-sequence lithium depletion factor $f_{\rm Li} = 24$, as shown in
Table~\ref{tab:results}.  This is a relatively large pre-main-sequence
lithium depletion, leaving only a factor of~$\sim 7$ to be accounted for
by main sequence effects (2) and~(3) above.  Note that the recent models of
\citet{BrunTZ99}
also appear to find solar pre-main-sequence lithium depletion by a factor of
order~10 (in good agreement with our results, considering the sensitivity of
pre-main-sequence lithium depletion to the input physics).

It is worth noting that, in the past,
pre-main-sequence lithium depletion in the Sun
has often been ignored or assumed to be negligible
(which is not the case, as our present models show).
Earlier models did in fact show relatively little pre-main-sequence solar
lithium depletion, e.g., a depletion factor of~3 was reported by
\citet{SBK93}.
These low lithium depletion factors may have been caused
partly by the use of older opacity tables in earlier models, but
a major contributing factor was neglect of gravitational settling (diffusion)
of helium and the heavy elements.  These earlier models thus had a
lower metallicity
during the pre-main-sequence stage, matching the present solar surface
metallicity, rather than being about 10\% higher (as diffusion models
indicate).  Because of the strong metallicity
dependence of pre-main-sequence lithium depletion, such models yielded
relatively small depletion factors.

Pre-main-sequence lithium depletion depends quite sensitively on the
structure of the solar models during that stage of evolution.  We give our
results in terms of the solar lithium depletion factor~$f_{\rm Li}$,
which is shown in Table~\ref{tab:results}.  The zoning did
not affect the lithium depletion factor significantly; nor did the
uncertainties in the solar age, luminosity, and radius.
Changes in the equation of state in the outermost regions can have a small
effect ($\lesssim 30$\%) on the lithium depletion factor:
the ``OPALeos-midT'' and
``OPALeos-lowT'' cases had lithium depletion factors of $f_{\rm Li} = 17$
and~19, respectively, as opposed to $f_{\rm Li} = 24$ for the
standard ``OPALeos-hiT'' case.

The pre-main-sequence lithium depletion is extremely sensitive to 
both low-temperature and high-temperature opacities; use of the
\citet{Shar92}
molecular opacities instead of the
\citet{AlexF94}
ones halved the lithium depletion factor ($f_{\rm Li} = 10$), as did
our ``low-$\kappa$'' test case ($f_{\rm Li} = 10$), while our
``high-$\kappa$'' test case nearly tripled it ($f_{\rm Li} = 71$).
There is also a relatively large sensitivity to
the uncertainty in the observed solar abundances.  A low $Z/X$ ratio
of~0.0203 (with ``$\kappa_{\rm OPAL:GS98\downarrow C-Ne}$'') yields
only two-thirds as much lithium depletion ($f_{\rm Li} = 15$), while a
high $Z/X$ ratio of~0.0257 (with ``$\kappa_{\rm OPAL:GS98\uparrow C-Ne}$'')
yields half again as much lithium depletion ($f_{\rm Li} = 35$).

Uncertainties in the diffusion constants can affect the lithium
depletion factor significantly, primarily due to the effect of the
different initial composition; our diffusion test cases have depletion
factors ranging from $f_{\rm Li} = 18$ to~33.

Except for the $\iso7{Li} + p$ rate, uncertainties in the nuclear rates
have almost no effect on the extent of lithium depletion.  For the
$\iso7{Li} + p$ rate, the $\pm 14$\% uncertainty quoted by the NACRE
compilation
\citep{Ang+99}
corresponds to an uncertainty of about $\pm 50$\% in the depletion factor
(i.e., a range in the from $f_{\rm Li} = 16$ to~38).

\paragraph{Solar beryllium abundance:}
The observed solar beryllium abundance is
$\log\,\varepsilon(\iso9{Be}) = 1.40 \pm 0.09$,
consistent with no depletion relative to the meteoritic value of
$\log\,\varepsilon(\iso9{Be}) = 1.42 \pm 0.04$.
The quoted uncertainties of these values imply that
solar beryllium cannot have been depleted by more than a
factor of~2 (3-$\sigma$ upper limit).
Our solar models all have negligible amounts of beryllium depletion,
of order~1\% --- in good agreement with these observational
beryllium results.  Other authors have shown that parameterized
rotational-mixing models which match the observed solar lithium depletion
yield relatively minor beryllium depletion, in agreement with observations.
For example, the rotational-mixing solar models of
\citet{BrunTZ99},
with total lithium depletion of a factor of order~100 (a factor
of~$\sim 10$ occurring on the main sequence), deplete beryllium by only
about~12\%.  Note that rotational-mixing models which ignore
pre-main-sequence lithium depletion (i.e., which over-estimate the
main-sequence lithium depletion by an order of magnitude) might be expected to
overestimate the beryllium depletion as well: for example,
\citet{Rich+96}
required main-sequence lithium depletion by a factor of~155 in their
solar models with rotational mixing, and found
that these models then implied beryllium depletion by a factor of~2.9.

\subsection{Solar neutrinos} \label{ssec:neutrinos}

We will not devote much space to the predicted solar neutrino values,
since it has long been concluded that matching the observed neutrino
capture rates requires not revised astrophysics but new neutrino physics,
e.g., Mikheyev-Smirnov-Wolfenstein (MSW) neutrino oscillation effects ---
even non-standard solar models (e.g., with core mixing or a low-metallicity
core) cannot simultaneously satisfy the neutrino observations and the
helioseismic constraints
\citep[see, e.g.,][]{BahBP98,Suz98,BasuPB00,BahPB01,WatanS01,TurckC+01a,%
TurckC+01b,Choub+01,Guzik+01}.
In Table~\ref{tab:results} we present the theoretically predicted neutrino
capture rates for the \iso{37}{Cl} and~\iso{71}{Ga} experiments,
and the predicted flux of \iso8B neutrinos.
As is normally obtained, our theoretical predicted neutrino rates are
much in excess of the observed values, i.e., 6.4 to 8.9 SNU is predicted
for the \iso{37}{Cl} experiment, as compared to the observed value of
$2.56 \pm 0.23$~SNU
%
%
\citep{Davis94,Clev+98},
and 127 to 141 SNU for the \iso{71}{Ga} experiments,
as compared to the observed value of $74.5 \pm 5.7$~SNU
%
%
\citep[combined value from SAGE and GALLEX+GNO:][]{Hamp+99,Abdur+99,%
Altmann+99,Gavrin01,Ferrari01}.
Likewise, the models predict \iso8B neutrino fluxes
of 4.4 to $6.3 \times 10^6 \; \rm cm^{-2}\;s^{-1}$,
as compared to the value of
$( 2.32 \pm 0.08 ) \times 10^6 \; \rm cm^{-2}\;s^{-1}$
implied by the Super-Kamiokande measurement
\citep{Fukuda+01}.

\section{Conclusion}

Helioseismic frequency observations
enable the adiabatic sound speed~$c$
and adiabatic index~$\gammaone$
to be inferred with an accuracy of a few parts in~$10^4$,
and the density~$\rho$ with an accuracy of a few parts in~$10^3$.
These quantities can also be computed on purely theoretical grounds.
It is important to understand the uncertainties in these theoretical
quantities (arising from uncertainties in the input data),
when comparing them to the values inferred from helioseismic
measurements.  These uncertainties in the theoretical standard solar
model are presented below.

{\bf (1)~Abundances of the elements:}
For the standard solar model,
we found that the largest impact on the
sound speed arises from the observational uncertainties in the
photospheric abundances of the elements.
The key elements for which
accurate meteoritic determinations are not available are C, N, O,
and~Ne, with uncertainties of~15\% (leading to an uncertainty of
order~10\% in the solar $Z/X$ ratio).  We determined that this
abundance uncertainty affects the sound speed profile in the solar model
at the level of 3 parts in~$10^3$.

{\bf (2)~OPAL opacities, $pp$ nuclear rate, and diffusion constants:}
The estimated 4\% uncertainty in the OPAL opacities, the $\sim 5$\%
uncertainty in the basic $pp$ nuclear reaction rate, the $\sim 15$\%
uncertainty in the
diffusion constants for the gravitational settling of helium, and
the $\sim 50$\% uncertainties in diffusion constants for the heavier elements,
all affect the sound speed at the level of 1 part in~$10^3$.

{\bf (3)~Solar radius and low-temperature equation of state:}
Different observational methods yield values for the solar radius
differing by as much as 7 parts in~$10^4$; this
leads to uncertainties of a few parts in~$10^3$ in the sound speed in the
solar convective envelope, but has negligible effect on the interior
(recall, however, that while the sound speed in the solar model is not
affected, there is a systematic effect on the ``observed''
{\it helioseismic\/} sound speed profile).
Uncertainties in the low-temperature equation of state ($\log\,T \lesssim 5.5$)
lead to uncertainties of order a part in~$10^3$ in both the sound speed and
the adiabatic index~$\gammaone$ in the convective envelope.

{\bf (4)~Rotational mixing and high-temperature equation of state:}
We did not explicitly consider the effects of rotational mixing or
uncertainties in the interior equation of state, but other investigators
have found these to yield uncertainties in the sound
speed of order a part in~$10^3$, as discussed in \S~\ref{ssec:profiles}.

{\bf (5)~Other sources of uncertainty:}
We found that other current uncertainties, namely, in the solar age and
luminosity, in nuclear rates other than the
$pp$~reaction and in the low-temperature molecular opacities,
have no significant effect on the
quantities that can be inferred from helioseismic observations
(although
some of these can have significant effects on neutrino fluxes and/or the
extent of pre-main-sequence lithium depletion).

{\bf (6)~Depth of envelope convection:}
Our reference standard solar model (with $Z/X = 0.0245$)
yielded a convective envelope position
$R_{ce} = 0.7135 \; R_\odot$, in excellent agreement with
the observed value of $0.713 \pm 0.001 \; R_\odot$, and was significantly
affected ($\pm 0.003 \; R_\odot$) only by uncertainties in~$Z/X$, opacities,
the $pp$~rate, and helium diffusion constants.

{\bf (7)~Envelope helium abundance:}
Our reference model yielded an envelope helium abundance
$Y_e = 0.2424$, in good agreement with the range of values
inferred from helioseismic observations (which we summarize as a
helioseismic $Y_e = 0.245 \pm 0.005$);
only extreme variations in~$Z/X$, opacities, or diffusion constants
yielded $Y_e$ variations as large as~0.005.

{\bf (8)~Pre-main-sequence lithium depletion:}
For the standard solar model, the predicted
pre-main-sequence lithium depletion is a factor of order~20
(an order of magnitude larger than that predicted by
earlier models that neglected gravitational settling and used
older opacities).
The lithium depletion factor can vary between $\sim 10$ and $\sim 40$ when
one varies the input physics, i.e., it is uncertain by a factor of~2.

{\bf (9)~Solar neutrinos:}
For the standard solar model, the predicted neutrino capture rate is
uncertain by~$\sim 30$\% for the \iso{37}{Cl} experiment and by~$\sim 3$\%
for the \iso{71}{Ga} experiments (not including uncertainties in the
capture cross sections), while the \iso8B neutrino flux is
uncertain by~$\sim 30$\%.

\acknowledgements

We are indebted to Prof.\ Marc H. Pinsonneault for helpful discussions on
diffusion and for providing us with his diffusion code.  We are grateful to
Prof.\ Sarbani Basu for discussions of helioseismology, and for providing
us with the current helioseismic reference model; we are also grateful to
Prof. Dimitri M. Mihalas, for providing us with his equation-of-state code.
We wish to thank Prof.\ Charles A. Barnes and Prof.\ Yuk L. Yung
for thoughtful discussions and encouragement.
We wish to acknowledge the support provided by Prof.\ Thomas A. Tombrello,
Chairman of the Division of Physics, Math, and Astronomy,
and Prof.\ Robert D. McKeown, Head of the W.~K.~Kellogg Radiation Laboratory.
One of us (I.-J.~S.) wishes to thank Alexandra R. Christy, her daughter,
and Prof.\ Robert F. Christy, her husband, for their
supportiveness, and Robert F. Christy
for critical analysis and helpful comments.  One of us (A.~I.~B.) wishes to
thank Prof.\ Peter G. Martin and Prof.\ J. Richard Bond for their support,
and M.~Elaine Boothroyd, his wife, for her patience and encouragement.
This work was supported by a grant \hbox{NAG5-7166} from the Sun-Earth
Connection Program of the Supporting Research and Technology
and Suborbital Program in Solar Physics of the National Aeronautics and
Space Administration, and by the National Science Foundation
grant \hbox{NSF-0071856} to the Kellogg Radiation Laboratory.


\clearpage


\begin{deluxetable}{lccccccccc@{\hspace{0pt}}c@{\hspace{0pt}}ccccc}

\tabletypesize{\scriptsize}

\rotate

\tablewidth{0pt}

\tablecaption{Characteristics of Our Solar
 Models\tablenotemark{a}\label{tab:results}}

\tablehead{
\colhead{} &
 \colhead{} & \colhead{} & \colhead{} & \colhead{} & 
 \colhead{$R_{ce}$} &
 \multicolumn{2}{c}{rms $\delta c/c$ for:} &
 \colhead{rms} &
 \multicolumn{3}{c}{relative rms} &
 \colhead{} & \colhead{} & \colhead{} & \colhead{} \\
\cline{7-8} \cline{10-12}
\colhead{Solar Model} &
 \colhead{$\alpha$} & \colhead{$Z_0$} & \colhead{$Y_0$} & \colhead{$Y_e$} &
 \colhead{($R_\odot$)} &
 \colhead{all-$r$} & \colhead{$< 0.6$} &
 \colhead{$\delta\rho/\rho$} &
 \colhead{vs.} & \colhead{$\Delta c/c$} & \colhead{$\Delta\rho/\rho$} &
 \colhead{$f_{\rm Li}$} & \colhead{$\Phi_{\rm Cl}$} &
 \colhead{$\Phi_{\rm Ga}$} & \colhead{$\Phi_{\rm B}$}
}

\startdata
%
1. Fine-zoned Reference\tablenotemark{b} &
 1.817 & .02030 & .2760 & .2424 & .7135 & .00133 & .00085 & .01698 &
  & \nodata & \nodata & 24.24 & 7.87 & 133.7 & 5.31 \\
\tableline
2. Fine-zoned, OPALeos-midT\tablenotemark{c} &
 1.803 & .02031 & .2760 & .2424 & .7133 & .00131 & .00085 & .01716 &
 \phn1 & .00029 & .00054 & 19.08 & 7.87 & 133.7 & 5.31 \\
%
3. Fine-zoned, OPALeos-lowT\tablenotemark{d} &
 1.814 & .02031 & .2760 & .2424 & .7134 & .00132 & .00085 & .01709 &
 \phn1 & .00013 & .00026 & 16.81 & 7.87 & 133.8 & 5.32 \\
\tableline
4. Coarse-zoned Reference\tablenotemark{b,e} &
 1.814 & .02030 & .2760 & .2424 & .7136 & .00140 & .00091 & .01772 &
 \phn1\tablenotemark{f} & .00008 & .00074 & 24.36 & 7.89 & 133.8 & 5.33 \\
\tableline
5. OPALeos-midT\tablenotemark{c} &
 1.800 & .02029 & .2760 & .2425 & .7135 & .00139 & .00092 & .01807 &
 \phn2\tablenotemark{f} & .00010 & .00092 & 19.74 & 7.90 & 134.0 & 5.34 \\
\tableline
6. $R_\odot = 695.78\;$Mm (R78) &
 1.815 & .02027 & .2760 & .2427 & .7134 & .00134 & .00089 & .01880 &
 \phn4 & .00045 & .00151 & 24.22 & 7.95 & 134.3 & 5.38 \\
%
7. $R_\odot = 695.508\;$Mm (R508) &
 1.819 & .02028 & .2760 & .2425 & .7129 & .00130 & .00092 & .01933 &
 \phn4 & .00098 & .00256 & 24.70 & 7.90 & 134.0 & 5.34 \\
%
8. R78, OPALeos-midT &
 1.803 & .02030 & .2760 & .2424 & .7131 & .00136 & .00085 & .01851 &
 \phn5\tablenotemark{g} & .00041 & .00097 & 19.95 & 7.89 & 133.8 & 5.33 \\
%
9. R508, OPALeos-midT &
 1.804 & .02027 & .2759 & .2427 & .7129 & .00148 & .00093 & .02001 &
 \phn5\tablenotemark{g} & .00099 & .00276 & 19.84 & 7.95 & 134.3 & 5.38 \\
\tableline
10. $L_\odot = L_{best}$\tablenotemark{h} &
 1.808 & .02029 & .2757 & .2421 & .7137 & .00143 & .00092 & .01739 &
 \phn4 & .00012 & .00053 & 23.95 & 7.73 & 132.8 & 5.21 \\
%
11. $L_\odot = L_{best} - 0.8$\% ($L_{low}$) &
 1.794 & .02030 & .2749 & .2416 & .7138 & .00145 & .00094 & .01681 &
 10\tablenotemark{g} & .00015 & .00107 & 23.39 & 7.38 & 130.3 & 4.94 \\
%
12. $L_\odot = L_{best} + 0.8$\% ($L_{high}$) &
 1.823 & .02028 & .2765 & .2429 & .7135 & .00140 & .00093 & .01839 &
 10\tablenotemark{g} & .00014 & .00133 & 24.62 & 8.16 & 135.7 & 5.54 \\
\tableline
13. $t_\odot = 4.5$ Gyr &
 1.801 & .02022 & .2768 & .2436 & .7149 & .00195 & .00148 & .02374 &
 \phn4 & .00064 & .00609 & 21.75 & 7.73 & 133.1 & 5.22 \\
%
14. $t_\odot = 4.7$ Gyr &
 1.826 & .02034 & .2752 & .2415 & .7124 & .00099 & .00057 & .01253 &
 \phn4 & .00056 & .00528 & 26.82 & 8.03 & 135.2 & 5.51 \\
\tableline
15. $Z_\kappa = Z_h$, CNO-interpolation &
 1.813 & .02030 & .2762 & .2425 & .7137 & .00144 & .00095 & .01818 &
 \phn4 & .00006 & .00047 & 24.16 & 7.90 & 134.0 & 5.36 \\
%
16. $Z_\kappa = Z_h$, CO$_{ex} = 0.0$\tablenotemark{i} &
 1.815 & .02030 & .2759 & .2424 & .7134 & .00135 & .00088 & .01735 &
 \phn4 & .00007 & .00039 & 24.55 & 7.86 & 133.7 & 5.30 \\
%
17. $Z_\kappa = Z_0$ (const-$Z_\kappa$) &
 1.823 & .02026 & .2759 & .2430 & .7073 & .00068 & .00046 & .01137 &
 \phn4 & .00098 & .00685 & 24.96 & 7.72 & 133.1 & 5.20 \\
%
18. $Z_\kappa = Z_0$, CO$_{ex} = Z - Z_\kappa$ &
 1.820 & .02027 & .2760 & .2428 & .7105 & .00092 & .00052 & .01364 &
 \phn4 & .00059 & .00428 & 24.71 & 7.79 & 133.5 & 5.25 \\
%
19. $Z_\kappa = Z$ (approx-$\kappa$) &
 1.812 & .02030 & .2764 & .2427 & .7138 & .00151 & .00103 & .01898 &
 \phn4 & .00016 & .00131 & 23.86 & 8.00 & 134.4 & 5.42 \\
%
20. $Z_\kappa = 0.9 \, Z$ (low-$\kappa$) &
 1.824 & .02056 & .2659 & .2331 & .7178 & .00313 & .00266 & .03376 &
 \phn4 & .00184 & .01619 & \phn9.72 & 7.08 & 129.8 & 4.70 \\
%
21. $Z_\kappa = 0.95 \, Z$ &
 1.819 & .02044 & .2713 & .2379 & .7156 & .00225 & .00178 & .02591 &
 \phn4 & .00094 & .00830 & 14.95 & 7.50 & 131.9 & 5.04 \\
%
22. $Z_\kappa = 1.05 \, Z$ &
 1.805 & .02016 & .2813 & .2474 & .7121 & .00094 & .00045 & .01315 &
 \phn4 & .00063 & .00471 & 39.97 & 8.58 & 137.4 & 5.87 \\
%
23. $Z_\kappa = 1.1 \, Z$ (high-$\kappa$) &
 1.797 & .02003 & .2860 & .2516 & .7106 & .00069 & .00055 & .00689 &
 \phn4 & .00126 & .01148 & 70.62 & 9.04 & 139.6 & 6.24 \\
%
24. $\kappa_{\rm LOAL85}$ at high T &
 1.882 & .02051 & .2688 & .2356 & .7193 & .00485 & .00380 & .03862 &
 \phn4 & .00359 & .02179 & 11.82 & 7.24 & 130.7 & 4.83 \\
%
25. $\kappa_{\rm Sharp}$ at low T &
 1.762 & .02027 & .2759 & .2426 & .7137 & .00144 & .00096 & .01820 &
 \phn4 & .00006 & .00050 & 12.06 & 7.94 & 134.2 & 5.37 \\
%
26. $\kappa_{\rm OPAL:GN93\uparrow C-Ne}$ (unmatched)\tablenotemark{j} &
 1.834 & .02045 & .2679 & .2356 & .7132 & .00159 & .00143 & .02330 &
 \phn4 & .00048 & .00565 & 15.12 & 7.03 & 130.2 & 4.74 \\
%
27. $\kappa_{\rm OPAL:GN93\downarrow C-Ne}$ (unmatched)\tablenotemark{j} &
 1.791 & .02009 & .2851 & .2504 & .7142 & .00140 & .00056 & .01219 &
 \phn4 & .00051 & .00572 & 46.66 & 8.91 & 138.8 & 6.14 \\
\tableline
28. $Z/X = .023$ (unmatched $\kappa$)\tablenotemark{j} &
 1.795 & .01923 & .2712 & .2376 & .7158 & .00227 & .00182 & .02653 &
 \phn4 & .00096 & .00891 & 15.16 & 7.29 & 130.5 & 4.89 \\
%
29. $Z/X = .023$, $\kappa_{\rm OPAL:GS98}$ &
 1.787 & .01911 & .2758 & .2419 & .7157 & .00192 & .00122 & .02125 &
 \phn4 & .00056 & .00374 & 22.60 & 7.76 & 132.8 & 5.26 \\
%
30. $Z/X = .0257$, $\kappa_{\rm OPAL:GS98\uparrow C-Ne}$ &
 1.840 & .02121 & .2768 & .2438 & .7113 & .00072 & .00046 & .01079 &
 29\tablenotemark{g} & .00152 & .01115 & 35.00 & 8.07 & 135.2 & 5.46 \\
%
31. $Z/X = .0203$, $\kappa_{\rm OPAL:GS98\downarrow C-Ne}$ &
 1.728 & .01700 & .2749 & .2396 & .7209 & .00366 & .00267 & .03375 &
 29\tablenotemark{g} & .00180 & .01276 & 14.81 & 7.41 & 130.2 & 5.04 \\
%
32. $Z/X = .0277$ (unmatched $\kappa$)\tablenotemark{j} &
 1.851 & .02251 & .2853 & .2517 & .7094 & .00091 & .00100 & .00315 &
 \phn4 & .00176 & .01704 & 77.94 & 9.20 & 141.3 & 6.30 \\
%
33. $Z/X = .0277$, $\kappa_{\rm OPAL:Gr84}$ &
 1.845 & .02255 & .2848 & .2510 & .7098 & .00076 & .00077 & .00851 &
 \phn4 & .00139 & .01095 & 52.25 & 9.87 & 145.6 & 6.76 \\
\tableline
34. high $pp$ rate &
 1.841 & .02014 & .2772 & .2447 & .7108 & .00074 & .00065 & .01042 &
 \phn4 & .00136 & .02614 & 24.20 & 6.61 & 127.2 & 4.37 \\
%
35. low $pp$ rate &
 1.805 & .02035 & .2756 & .2417 & .7146 & .00185 & .00141 & .02683 &
 \phn4 & .00051 & .00936 & 24.32 & 8.44 & 136.7 & 5.75 \\
%
36. high $\iso3{He}+\alpha$ rate &
 1.817 & .02026 & .2763 & .2430 & .7132 & .00127 & .00078 & .01452 &
 \phn4 & .00018 & .00349 & 24.03 & 8.84 & 140.1 & 6.04 \\
%
37. low $\iso3{He}+\alpha$ rate &
 1.810 & .02031 & .2757 & .2421 & .7140 & .00157 & .00111 & .02186 &
 \phn4 & .00021 & .00432 & 24.43 & 6.91 & 127.7 & 4.60 \\
%
38. high $\iso3{He}+\iso3{He}$ rate &
 1.813 & .02028 & .2759 & .2425 & .7137 & .00146 & .00098 & .01883 &
 \phn4 & .00008 & .00115 & 24.06 & 7.79 & 133.3 & 5.26 \\
%
39. low $\iso3{He}+\iso3{He}$ rate &
 1.815 & .02028 & .2761 & .2426 & .7135 & .00138 & .00090 & .01735 &
 \phn4 & .00008 & .00049 & 24.22 & 8.09 & 135.2 & 5.48 \\
%
40. high $p+\iso{14}N$ rate &
 1.816 & .02029 & .2762 & .2426 & .7134 & .00133 & .00084 & .01616 &
 \phn4 & .00008 & .00168 & 24.30 & 8.15 & 136.4 & 5.36 \\
%
41. low $p+\iso{14}N$ rate &
 1.813 & .02029 & .2758 & .2425 & .7137 & .00149 & .00102 & .01993 &
 \phn4 & .00024 & .00238 & 24.24 & 7.78 & 131.7 & 5.35 \\
%
42. high $p+\iso7{Be}$ rate &
 1.814 & .02030 & .2760 & .2424 & .7136 & .00140 & .00091 & .01771 &
 \phn4 & .00001 & .00004 & 24.36 & 8.50 & 135.2 & 5.89 \\
%
43. low $p+\iso7{Be}$ rate &
 1.814 & .02030 & .2760 & .2424 & .7136 & .00140 & .00091 & .01772 &
 \phn4 & .00001 & .00004 & 24.36 & 7.26 & 132.5 & 4.77 \\
%
44. high $p+\iso7{Li}$ rate &
 1.814 & .02030 & .2760 & .2424 & .7136 & .00140 & .00091 & .01771 &
 \phn4 & .00001 & .00006 & 38.07 & 7.89 & 133.8 & 5.33 \\
%
45. low $p+\iso7{Li}$ rate &
 1.814 & .02030 & .2760 & .2424 & .7136 & .00139 & .00091 & .01772 &
 \phn4 & .00001 & .00005 & 15.69 & 7.89 & 133.8 & 5.33 \\
%
46. CF88 nuclear rates &
 1.832 & .02018 & .2767 & .2440 & .7117 & .00077 & .00035 & .00500 &
 \phn4 & .00090 & .01709 & 16.28 & 7.85 & 130.4 & 5.46 \\
\tableline
47. 20\% low $D_{\rm He}$ &
 1.781 & .02018 & .2760 & .2485 & .7169 & .00217 & .00162 & .02439 &
 \phn4 & .00084 & .00682 & 20.15 & 7.71 & 132.9 & 5.20 \\
%
48. 20\% high $D_{\rm He}$ &
 1.846 & .02038 & .2760 & .2368 & .7106 & .00080 & .00046 & .01171 &
 \phn4 & .00081 & .00635 & 29.03 & 8.12 & 135.2 & 5.51 \\
%
49. 40\% low $D_Z$ &
 1.824 & .01949 & .2725 & .2385 & .7128 & .00165 & .00130 & .02187 &
 \phn4 & .00038 & .00420 & 18.43 & 7.42 & 131.2 & 4.98 \\
%
50. 40\% high $D_Z$ &
 1.804 & .02110 & .2795 & .2463 & .7145 & .00129 & .00059 & .01412 &
 \phn4 & .00033 & .00369 & 32.46 & 8.42 & 136.9 & 5.72 \\
%
\enddata

\tablenotetext{a}{ Mixing length parameter~$\alpha$, pre-solar
 metallicity~$Z_0$ and helium mass fraction~$Y_0$, present envelope
 helium abundance~$Y_e$, position~$R_{ce}$ of the base of envelope
 convection, rms fractional sound speed and density differences relative
 to the Sun's inferred helioseismic profiles and relative to the reference
 standard solar model, pre-main-sequence lithium depletion
 factor~$f_{\rm Li}$, predicted capture rates (in SNU) $\Phi_{\rm Cl}$ and
 $\Phi_{\rm Ga}$ for chlorine and gallium experiments, respectively, and
 predicted flux~$\Phi_{\rm B}$ of \iso8B neutrinos (in units of
 $10^6$~cm~s$^{-1}$).}

\tablenotetext{b}{ Reference standard solar model: OPAL EOS at
 $\log\,\rho \gtrsim -1.5$, MHD EOS at $\log\,\rho \lesssim -2$,
 high-T opacities $\kappa_{\rm OPAL:GN93}$ interpolated in
 $Z_\kappa = Z_h \equiv Z_0 \, [ \sum_{heavy} X_i ] / [ \sum_{heavy} (X_i)_0 ]$
 as well as in ``excess'' C and~O (such that
 ${\rm C}_{ex} + {\rm O}_{ex} \equiv {\rm CO}_{ex} = Z - Z_\kappa$),
 low-T opacities $\kappa_{\rm Alexander}$, NACRE nuclear rates,
 gravitational settling of He and heavy elements, $Z/X = 0.0245$,
 $L_\odot = 3.854 \times 10^{33}$~erg~s$^{-1}$,
 $R_\odot = 695.98$~Mm, and $t_\odot = 4.6$~Gyr; both fine-zoned and
 coarse-zoned cases were computed.  Variant models have the same input
 values, except as specified in the first column.}

\tablenotetext{c}{ OPALeos-midT: use OPAL EOS at $\log\,T \gtrsim 4.0$,
 MHD EOS at $\log\,T \lesssim 3.9$.}

\tablenotetext{d}{ OPALeos-lowT: use OPAL EOS at $\log\,T \gtrsim 3.75$,
 MHD EOS at $\log\,T \lesssim 3.7$ (on the main sequence, MHD EOS
 is only used outside the photosphere).}

\tablenotetext{e}{ This reference standard solar model and all subsequent
 models in the table use the coarse zoning.}

\tablenotetext{f}{ For these cases, the relative rms values
 compare the coarse-zoned cases to the corresponding fine-zoned cases.}

\tablenotetext{g}{ For these cases, the relative rms values
 compare the variant models with the case relative to which
 the parameter variation was performed.}

\tablenotetext{h}{ Most recent solar luminosity value of
 $L_\odot = L_{best} \equiv 3.842 \times 10^{33}$ erg s$^{-1}$.}

\tablenotetext{i}{ OPAL opacities not interpolated in ``excess'' C and~O,
 i.e., opacities computed as if the CNO element abundance profiles in the
 Sun were always in the same proportion to the Fe~profile ($Z_k = Z_h \ne Z$, 
 but no correction term: ${\rm CO}_{ex} = 0.0$).}

\tablenotetext{j}{ ``Unmatched $\kappa$'' means that $Z/X$ has been varied
 relative to the recommended value of the given mixture (for which the OPAL
 opacities had been computed).}

\end{deluxetable}


\clearpage


\begin{deluxetable}{ccl}

\tablewidth{0pt}

\tablecaption{Helioseismically Measured Present Solar Envelope Helium
 Abundance $Y_e$\label{tab:Yenv}}

\tablehead{
\colhead{$Y_e$ Using OPAL Eos} & \colhead{$Y_e$ Using MHD Eos} &
 \colhead{Reference}
}

\startdata
\nodata           & $0.242 \pm 0.003$ & 1 \\
$0.249 \pm 0.001$ & $\approx 0.25$    & 2 \\
$\approx 0.23$\tablenotemark{a} & $\approx 0.25$\tablenotemark{a} & 3 \\
$0.248 \pm 0.001$ & $0.232 \pm 0.006$ & 4 \\
$0.248 \pm 0.001$ & $\approx 0.252$   & 5 \\
$0.248 \pm 0.002$ & $\approx 0.242$   & 6 \\
$\approx 0.226$\tablenotemark{b} & \nodata & 7 \\
$0.2539\pm0.0005$ & $0.2457\pm0.0005$ & 8 \\
\enddata

\tablenotetext{a}{ Obtained via an entropy calibration.
}

\tablenotetext{b}{ Obtained from a fit to a solar sound speed profile,
not a direct helioseismic inversion.
}

\tablerefs{(1)~\citealt{PerC94},
(2)~\citealt{BasuA95},
(3)~\citealt{BatA97},
(4)~\citealt{Kos97},
(5)~\citealt{Basu98},
(6)~\citealt{Rich+98},
(7)~\citealt{ShiHT99},
(8)~\citealt{DiMauro+02}
}

\end{deluxetable}


\end{document}